\documentclass[review]{elsarticle}

\usepackage[a4paper]{geometry}

\makeatletter
\def\ps@pprintTitle{%
    \let\@oddhead\@empty
    \let\@evenhead\@empty
    \let\@oddfoot\@empty
    \let\@evenfoot\@oddfoot
}
\makeatother

\usepackage{multirow}
\usepackage{booktabs}
\usepackage{siunitx}
\usepackage{tikz}
\usepackage{amsmath, amssymb, mathtools}
\usepackage{subcaption}
\usepackage{algorithm, algpseudocode}
\usepackage{url}
\usepackage{multicol}

\newcommand{\reviewchange}[1]{{\color{black}#1}\color{black}}

\DeclareMathOperator{\asinh}{sinh^{-1}}

\DeclareMathOperator*{\argmin}{arg\,min}
\DeclareMathOperator*{\argmax}{arg\,max}

\newcommand{\dx}[1]{\mathrm{d}#1}
\newcommand{\moment}[1]{\left\langle#1\right\rangle}
\newcommand{\norm}[2]{\left\lvert\left\lvert #2\right\rvert\right\rvert_#1}

\newcommand{\xvec}{\boldsymbol{x}}
\newcommand{\xivec}{\boldsymbol{\xi}}
\newcommand{\alphavec}{\boldsymbol{\alpha}}
\newcommand{\yvec}{\boldsymbol{y}}
\newcommand{\cvec}{\boldsymbol{c}}
\newcommand{\rvec}{\boldsymbol{r}}

\newcommand{\daedalus}{DAE$\delta$ALUS}
\newcommand{\isodar}{IsoDAR}

\usetikzlibrary{mindmap,trees, shapes, arrows}
\pgfdeclarelayer{background}
\usetikzlibrary{snakes,backgrounds, positioning, arrows.meta}

\newcommand{\Figref}[1]{Fig.~\ref{#1}}
\newcommand{\Eqref}[1]{Eq.~\eqref{#1}}
\newcommand{\Secref}[1]{Sec.~\ref{#1}}
\newcommand{\Tabref}[1]{Tab.~\ref{#1}}
\newcommand{\Algref}[1]{Alg.~\ref{#1}}

\newcommand{\uqtk}{UQTk}
\newcommand{\opal}{OPAL}

\newcommand{\ringcyc}{PSI Ring}
\newcommand{\injII}{PSI Injector II}

\begin{document}
    
    \begin{frontmatter}
    \title{Global Sensitivity Analysis on Numerical Solver Parameters of Particle-In-Cell Models
    in Particle Accelerator Systems}

    \author[]{Matthias Frey\corref{cor}} 
    \ead{matthias.frey@psi.ch}

    \author[]{Andreas Adelmann}
    \ead{andreas.adelmann@psi.ch}

    \address{Paul Scherrer Institut, CH-5232 Villigen, Switzerland} 

    \cortext[cor]{Corresponding author}

    \begin{abstract}
    Every computer model depends on numerical input parameters that are chosen according to mostly conservative
    but rigorous numerical or empirical estimates. 
    These parameters could for example be the step size for time integrators, a seed for pseudo-random number generators,
    a threshold or the number of grid points to discretize a computational domain. In case a numerical model is enhanced
    with new algorithms and modelling techniques the numerical influence on the quantities of interest, the running time
    as well as the accuracy is often initially unknown. \\
    Usually parameters are chosen on a trial-and-error basis neglecting the computational cost versus accuracy aspects. As
    a consequence the cost per simulation might be unnecessarily high which wastes computing resources.
    Hence, it is essential to identify the most critical numerical
    parameters and to analyze systematically their effect on the result in order to minimize the time-to-solution without losing
    significantly on
    accuracy. Relevant parameters are identified by global sensitivity studies where Sobol' indices are common measures. These
    sensitivities are obtained by surrogate models based on polynomial chaos expansion.\\
    In this paper, we first introduce the general methods for uncertainty quantification. We then demonstrate their use
    on numerical solver parameters to reduce the computational costs and discuss further model improvements based
    on the sensitivity analysis.
    The sensitivities are evaluated
    for neighbouring bunch simulations of the existing \injII{} and \ringcyc{} as well as the proposed \daedalus{} Injector
    cyclotron and simulations of the rf electron gun of the Argonne Wakefield Accelerator.
    \end{abstract}


    \end{frontmatter}
    
\section{Introduction}
Numerical models in scientific research disciplines are usually extremely complex and
computationally intensive. A common feature to all is the dependency on
numerical model parameters that do not represent an actual property of the
underlying scientific problem.
They are either prescribed in the source code and, thus, hidden to the user or
can be chosen at runtime. Examples are the seed for pseudo random number
generators, an error threshold in a convergence criterion, the number of grid
points in mesh-based models or the step size in time integrators. Latter two
are often chosen to satisfy memory constraints or time limits. 
When applying new algorithms and modelling techniques the sensitivity of such input
values on the response is usually studied by varying only a single parameter. While
this captures the main influence of the tested parameter, possible correlations with
other parameters are missed. A remedy is the evaluation of Sobol' indices
\cite{SOBOL2001271} which are variance-based global sensitivity measures to express
both individual and correlated parameter influences. Instead of Monte Carlo estimates,
these quantifiers can easily be obtained by surrogate models based on polynomial chaos
expansion (PCE). Uncertainty quantification (UQ) based on PCE is generally used in many areas
of scientific computing and modelling \cite{doi:10.1146/annurev.fluid.010908.165248, doi:10.1088/1364-7830/8/3/010,
doi:10.1002/cnm.2755, doi:10.1137/16M1061928} and several frameworks exist such as
\cite{doi:10.1061/9780784413609.257, FEINBERG201546, dakota}. The coefficients of the truncated PCE that are
required to determine
Sobol' indices are mostly computed using the projection or regression method \cite{SUDRET2008964}. Since some
numerical parameters are limited to integers, the former method is not applicable.

In this paper we study the sensitivity of adaptive mesh refinement (AMR) and multi bunch
\cite{PhysRevSTAB.9.064402, PhysRevSTAB.13.064201} parameters in Particle-In-Cell (PIC)
simulations of high intensity cyclotrons where we use the new AMR capabilities of \opal{}
(Object Oriented Parallel Accelerator Library) \cite{2019arXiv190506654A} as presented in \cite{FREY2019106912}. We further
explore the sensitivity of a rf electron gun model with respect to the number of macro particles, the energy binning and the
time step.
The sensivitities are evaluated using ordinary least squares and Bayesian compressive sensing (BCS) \cite{BCS_4524050, BCS_5256324}.
Both methods are part of the uncertainty quantification toolkit \uqtk{} \cite{doi:10.1137/S1064827503427741, Debusschere2017} \reviewchange{(version 3.0.4)}. The results
are cross-checked using Chaospy \cite{FEINBERG201546} \reviewchange{(version 3.0.5)}{} together with the orthogonal matching pursuit \cite{258082, 342465}
regression model of the Python machine learning library scikit-learn \cite{scikit-learn, sklearn_api} \reviewchange{(version 0.21.2)}.

Although we apply UQ on numerical solver parameters, it is a general method
used for example in  \cite{doi:10.1137/16M1061928} to evaluate the sensitivities and to predict the quantities of interest due to uncertainties in physical parameters.

The paper is organised as follows: In \Secref{sec:application} we elaborate the physical applications and their
numerical modelling. An introduction to UQ is given in \Secref{sec:uq}. The experimental setup and
its results are presented in \Secref{sec:experiment_design} and \Secref{sec:results}, respectively. In \Secref{sec:conclusions}
are the final conclusions.

\section{Applications}

\subsection{High Intensity Cyclotrons}
\label{sec:application}
Cyclotrons are circular machines that accelerate charged particles (e.g. protons)
or ions (e.g. $H_{2}^{+}$). Depending on the particle species and delivered energy, these machines find different
applications ranging from isotope production \cite{MORRISSEY200390, Alonso2019} and neutron
spallation \cite{FISCHER19971202} to cancer treatment \cite{MCCARTHY199735, WEBER2005401}.
An example that provides a beam for neutron spallation is the High Intensity Proton Accelerator (HIPA) facility
at Paul Scherrer Institut (PSI) consisting of two cyclotrons, i.e. the \injII{} and \ringcyc. At a frequency of \SI{50.65}{MHz}
\SI{10}{mA} DC (direct current) proton bunches at \SI{870}{keV} are injected into \injII{} in which they are collimated
to approximately \SI{2.2}{mA} and accelerated up to \SI{72}{MeV}
($\sim\SI[round-mode=places, round-precision=0]{37.0751605}{\percent}$ speed of
light), before being transported to the \ringcyc{} where they reach a kinetic energy of \SI{590}{MeV}
($\sim\SI[round-mode=places, round-precision=0]{78.935030854}{\percent}$ speed of light) at extraction.

Another example is the planned facility of the \daedalus{} and \isodar{} (Isotope Decay At Rest) experiments for neutrino
oscillation and CP violation. It consists of two cyclotrons where the \daedalus{} Injector Cyclotron (DIC) \cite{doi:10.1063/1.4802375, doi:10.1063/1.4826879} is the first acceleration stage delivering a \SI{60}{MeV/amu} $H_{2}^{+}$ beam to the Superconducting Ring Cyclotron (DSRC) with
extraction energy of \SI{800}{MeV/amu}.

\begin{figure}[!ht]
    \centering
    \begin{subfigure}[t]{0.48\textwidth}
    \centering
    \includegraphics[width=1.0\textwidth]{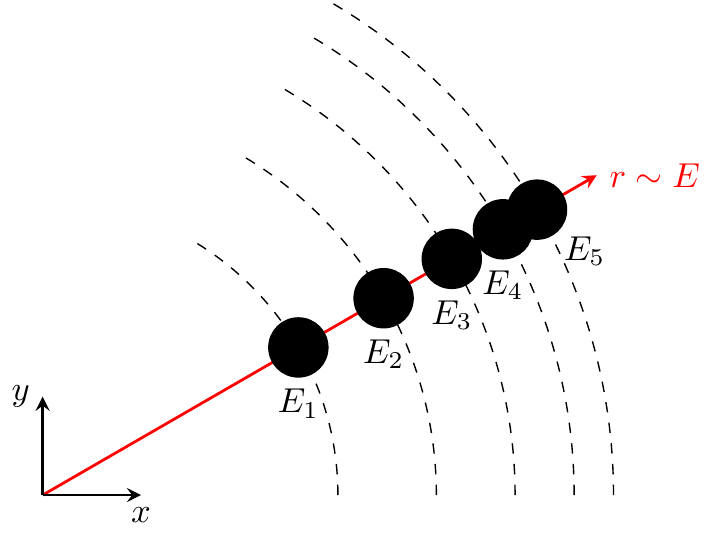}
    \caption{Five bunches evolving radially on axis due to the isochronism of the cyclotron. The origin
    of the coordinate system denotes the center of the machine. The orbit radius $r$ of each bunch is proportional to
    its energy $E_{1} < E_{2} < \dots < E_{5}.$}
    \label{fig:cyc_sketch}
    \end{subfigure}
    \hfill
    \begin{subfigure}[t]{0.48\textwidth}
        \centering
        \includegraphics[width=0.5\textwidth]{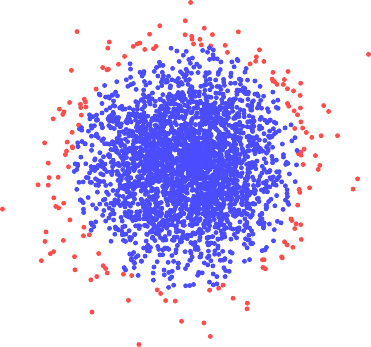}
        \caption{Separation of a particle bunch into core (blue) and halo (red) particles. In the \ringcyc{} the overall
        loss is on the order of $10^{-4}$ which corresponds to a beam intensity of about $\SI{0.2}{\micro A}$, i.e.
        all particles outside of approximately $3\sigma$ of a Gaussian distribution with standard deviation $\sigma$.}
        \label{fig:bunch_sketch}
    \end{subfigure}
    \caption{Sketch of neighbouring bunches (left) in the context of isochronous cyclotrons and
    characterization of a single bunch (right).}
\end{figure}

Since cyclotrons are isochronous, i.e. the magnetic field
is increased radially in order to keep the orbital frequency constant, i.e. the revolution time per turn is energy-independent
and, thus, the bunches lie radially on axis. A sketch showing five bunches denoted as circles on adjacent turns
is depicted in \Figref{fig:cyc_sketch}. As shown in \cite{PhysRevSTAB.13.064201}, a small turn separation causes
interactions between neighbouring bunches which yields to more halo particles (cf. \Figref{fig:bunch_sketch}).
In order to resolve these effects, the open source beam dynamics code \opal{} \cite{2019arXiv190506654A} got recently enhanced
with AMR capabilities \cite{FREY2019106912} which adds more complexity to the numerical model. The influence of the AMR solver
parameter settings on the statistical measures of the particle bunches is yet unknown and a too conservative AMR regrid frequency
worsens the time-to-solution. Furthermore, the applied energy binnnig technique
\cite{PhysRevSTAB.9.064402} to fulfill the electrostatic assumption (cf. \Secref{sec:numerical_model}) increases the computational
costs considerably. Hence, the goal of this study is to quantify the impact of AMR solver and energy binning
parameters in order to improve further computational investigations of these bunch interactions.

\subsubsection{Numerical Model}
\label{sec:numerical_model}
The numerical model of neighbouring bunch simulations in \opal{}, as presented in \cite{PhysRevSTAB.13.064201}, is based on
\cite{PhysRevSTAB.9.064402}. Due to the energy difference of the particle bunches on neighbouring turns a single transformation into
the particle rest frame does not fully satisfy the requirements of the electrostatic assumption to solve Poisson's equation. Instead,
each of the $N$ macro particles is assigned to an energy bin
$b$ due to its momentum $\beta\gamma$ according to
\begin{equation}
    b = \Bigg\lfloor\frac{\asinh{\left(\beta\gamma\right)} -
                           \asinh{\left(\min_{i=\{1,N\}}(\beta\gamma)_i\right)}}
                          {\eta}\Bigg\rfloor
    \label{eq:binning}
\end{equation}
where the binning parameter $\eta$ is a measure of the energy spread.
In each time step the force on a particle exerted by all others is the sum
of the electric field contributions of each energy bin $b$ evaluated in the appropriate rest frame of
the particles obtained by a Lorentz transform with the proper relativistic factor
$\gamma_b$. The algorithm is summarised in \Algref{alg:space_charge_calculation}. The computation of the electric
field of an energy bin involves only particles of that bin, thus, the charge deposition applies only on a subset of particles
$M\subset{\{1,\dots,N\}}$. However, the field on the mesh is interpolated and applied to all $N$ particles.

\begin{algorithm}[!ht]
    \small
    \begin{algorithmic}[1]
        \State ${\bf E}_{n} \gets {\bf 0}\quad \forall n\in\{1,\dots,N\}$
        \Comment{Electric field at particle location}

        \For{$b\in\{1,\dots,B\}$}\Comment{Loop over energy bins}
            \State${\bf \tilde{x}}_n\gets$\Call{LorentzTransform}{{${\bf x}_n, \gamma_b$}}$\quad \forall n\in\{1,\dots,N\}$
            \State$\tilde{\rho}_{i,j,k}\gets$\Call{DepositCharge}{${\bf \tilde{x}}_m, q_m$}
            $\quad \forall m\in M\subset{\{1,\dots,N\}}$
            \Comment{Interpolate charge onto mesh}
            \State${\bf \tilde{E}}_{i,j,k}^b \gets$\Call{PoissonSolve}{$\tilde{\rho}_{i,j,k}$}
            \State${\bf \tilde{E}}_{n}^{b}\gets$
            \Call{GatherEfield}{${\bf \tilde{E}}_{i,j,k}^b, {\bf \tilde{x}}_n$}$\quad \forall n\in\{1,\dots,N\}$
            \Comment{Get field at particle location}
            \State${\bf E}_{n}^{b}\gets$\Call{BackLorentzTransform}{{${\bf \tilde{E}}_n^b,  \gamma_b$}}
            $\quad \forall n\in\{1,\dots,N\}$
            \State${\bf x}_n\gets$\Call{BackLorentzTransform}{{${\bf \tilde{x}}_n, \gamma_b$}}$\quad \forall n\in\{1,\dots,N\}$
            \State ${\bf E}_{n}\gets {\bf E}_{n} + {\bf E}_{n}^{b}\quad \forall n\in\{1,\dots,N\}$
            \Comment{Add field contribution}
        \EndFor
    \end{algorithmic}
    \caption{Electrostatic Particle-In-Cell with $B$ energy bins and $N$ macro particles.}
    \label{alg:space_charge_calculation}
\end{algorithm}

\subsubsection{RF Electron Gun Model}
To study the effect of energy binning we further use the example of the Argonne Wakefield Accelerator (AWA)
\cite{gai_power_jing_2012, JING201872, Conde:IPAC2017-WEPAB132} facility, an experiment setup for beam physics studies and
accelerator technology developments. The facility is equipped with a photocathode rf electron gun that emits high intensity electron beams at
high accelerating gradients ($\gg\SI{1}{MV/m}$).
Due to the high gradients the electrostatic approximation is invalidated and, hence, energy binning is necessary. In \opal, we model
the particle emission by
\begin{equation*}
    \begin{aligned}
        p_{total} &= \sqrt{\left(\frac{E_{kin}}{mc^2}+1\right)^2-1} \\
        p_x &= p_{total}\sin\left(\varphi\right)\cos\left(\theta\right) \\
        p_y &= p_{total}\sin\left(\varphi\right)\sin\left(\theta\right) \\
        p_z &= p_{total}\left|\cos\left(\varphi\right)\right| \\
        \varphi &= 2\cos^{-1}\left(\sqrt{x}\right)
    \end{aligned}
\end{equation*}
with $x\in[0,1]$ and $\theta\in[0,\pi]$ \reviewchange{uniformly}{} randomly sampled
\cite{2019arXiv190506654A}.

\subsection{Quantities of interest}
In accelerator physics interesting quantities of interest (QoI) also denoted as
observables, of the co-moving frame are the rms beam size
\begin{equation*}
\sigma_\omega = \sqrt{\langle \omega^2\rangle}\quad\forall \omega = x, y, z
\end{equation*}
and the projected emittance
\begin{equation}
\varepsilon_\omega = \sqrt{\langle\omega^2\rangle
                           \langle p_\omega^2\rangle - \langle\omega p_\omega\rangle^2}
    \label{eq:proj_emit}
\end{equation}
that describes the phase space volume per dimension. The bracket $\langle\cdot\rangle$ represents the moment.
In order to quantify halo (cf. \Figref{fig:bunch_sketch}), i.e.
the tails of a particle distribution, we use two statistical definitions for bunched beams
by \cite{Wangler, PhysRevSTAB.5.124202}, the spatial-profile parameter
\begin{equation}
    h_\omega = \frac{\langle\omega^4\rangle}{\sigma_\omega^4} - \frac{15}{7}
    \label{eq:profile_parameter}
\end{equation}
and the phase-space halo parameter
\begin{equation}
    H_\omega = \frac{\sqrt{3}}{2} \frac{\sqrt{I_4^\omega}}{\varepsilon_\omega^2} - \frac{15}{7}
    \label{eq:halo_parameter}
\end{equation}
with \Eqref{eq:proj_emit} and fourth order invariant (cf. also \cite{doi:10.1063/1.37804})
\begin{equation*}
            I_4^\omega = \left\langle \omega^4\right\rangle\left\langle p_\omega^4\right\rangle
                    + 3\left\langle \omega^2 p_\omega^2\right\rangle^2
                    - 4\left\langle \omega p_\omega^3\right\rangle\left\langle \omega^3 p_\omega\right\rangle.
\end{equation*}
In case the bunch has uniform density \Eqref{eq:profile_parameter} is zero due to the constant $15/7$. An important
quantity of interest in the rf electron gun model is the energy spread
\begin{equation}
\Delta E \propto \sqrt{\langle p_{z}^2\rangle}. 
\label{eq:energy_spread}
\end{equation}

\section{Non-Intrusive Uncertainty Quantification}
\label{sec:uq}
\reviewchange{Simulations of physical phenomena usually rely on measured input data. Depending on the accuracy of the measurement and the model
sensitivity, the response may vary significantly. UQ introduces methods to quantify this variability in order to estimate the
reliability of the obtained results. UQ distinguishes two approaches which are called intrusive and non-intrusive UQ}.
In contrast to intrusive UQ, non-intrusive UQ uses the computational model as a black box. In this paper we only give a short overview
following the description and notation of \cite{doi:10.1137/16M1061928, SUDRET2008964, SOBOL2001271, Sargsyan_2014, Sargsyan2017}. A
detailed introduction to UQ in general is found in e.g. \cite{HandbookOfUQ, Sullivan2015}.

\reviewchange{In \Secref{sec:3_1}, the surrogate model based on the polynomial chaos expansion (PCE) is explained in general. The sections
\ref{sec:uq_projection_method} to \ref{sec:BCS} describe methods to obtain the coefficients of the expansion where special focus is
given on the methods applied in this paper. The definition of the Sobol' indices and their computation with the PCE is given in
\Secref{sec:Sensitivity_Analysis}. In order to check the error bounds of the estimated sensitivities, we use the bootstrap
method explained in \Secref{sec:ci_bootstrap}}.

\subsection{Surrogate Model based on Polynomial Chaos Expansion}
\label{sec:3_1}
The PC-decomposition originates from \cite{10.2307/2371268}, where a random variable of a Gaussian distribution is represented as a
series of multivariate Hermite polynomials of increasing order.
\reviewchange{And as stated by the Cameron-Martin theorem \cite{10.2307/1969178},  any functional in $L_2$ space can be represented in
a series of Fourier-Hermite functionals. Later, this method was rediscovered and applied by \cite{SpanosGhanem} to a stochastic
process and then generalized by \cite{doi:10.1137/S1064827501387826} to other probability measures and their corresponding orthogonal
polynomials.}

\reviewchange{Let a multivariate polynomial $\Psi_{\alphavec_{i}}(\xivec)$ of dimension $d\in\mathbb{N}\setminus\{0\}$
and multiindex $\alphavec_{i} = (\alpha_1, \alpha_2, \dots, \alpha_d)\in\mathbb{N}^d$ be defined by
\begin{equation*}
\Psi_{\alphavec_{i}}(\xivec) = \prod_{j=1}^{d}\psi_{\alpha_j}(\xi_j)
\end{equation*}
with orthogonal univariate polynomials $\{\psi_{\alpha_j}\}_{j=1}^{d}$. The response of a model $m(\xvec)$ with random input vector
$\xvec\in\Omega_1\times\dots\times\Omega_d$, where $\Omega_{j}~\forall j = \{1,\dots,d\}$ denotes the sample space of
the $j$-th random variable, can then be represented as
\begin{equation}
    m(\xvec) = \sum_{i=0}^{\infty}c_{\alphavec_i}\Psi_{\alphavec_i}(\mathcal{T}(\xvec))
    \label{eq:uq_no_approx}
\end{equation}}
where the basis of a random input component is determined by its probability distribution (cf.
\Tabref{tab:uq_distribution_polynomial})
and $\mathcal{T}: \xvec\mapsto\xivec$ denotes an isoprobabilistic transform. In case of dependent input components, for example,
$\mathcal{T}$ represents the Rosenblatt transform \cite{10.2307/2236692} that yields independent random variables. Another
method for dependent variables presented in \cite{JAKEMAN2019643} applies the Gram-Schmidt orthogonalization.

Under the assumption of only independent input variables, the transform $\mathcal{T}$ reduces to a simple linear mapping of every
component of $\xvec$ onto the defined interval of the corresponding univariate polynomial, e.g. $\xi_j\in [-1, 1]$ for
Legendre polynomials.
In numerical computations the sum in \Eqref{eq:uq_no_approx} is truncated at some polynomial degree $p$, hence the expansion is only an
approximation of the exact model $m(\xvec)$, i.e.
\begin{equation}
    \hat{m}(\xvec) = \sum_{i=0}^{P-1}c_{\alphavec_i}\Psi_{\alphavec_i}(\mathcal{T}(\xvec)).
    \label{eq:uq_approx}
\end{equation}
The truncation scheme is not clearly defined. A common rule, which is also used here, is the so-called total order truncation that
keeps all multiindices $\alphavec$ for which $\norm{1}{\alphavec} \le p$. This yields a number of
\begin{equation}
    P = \frac{(p+d)!}{p!d!}
    \label{eq:num_TO_indices}
\end{equation}
multiindices. Three other schemes are explained in \cite{Sargsyan_2014}.
In the next sections we describe methods to compute the coefficients $c_{\alphavec_i}$ of \Eqref{eq:uq_approx}.

\begin{table}
    \centering
    \begin{tabular}{lll}
        \toprule
        PDF of $\xi_j$ & Polynomial Basis       & Support \\
                       & \reviewchange{$\{\psi_{\alpha_{j}}(\xi_j)\}$}    & $\Omega_j$ \\
        \midrule
        Gaussian    & Hermite   & $]-\infty, \infty[$ \\
        Gamma       & Laguerre  & $[0, \infty[$ \\
        Uniform     & Legendre  & $[a, b]$ with $a,b\in\mathbb{R}$\\
        \bottomrule
    \end{tabular}
    \caption{Examples of the Wiener-Askey polynomial chaos of random variables $\xi_j$ with appropriate probability density function
    (PDF) \cite{doi:10.1137/S1064827501387826}.}
    \label{tab:uq_distribution_polynomial}
\end{table}

\subsection{Projection Method}
\label{sec:uq_projection_method}
The (spectral) projection method computes the coefficients of
\Eqref{eq:uq_approx} making use of the orthogonality of the basis functions, i.e.
\reviewchange{$\left\langle\Psi_{\alphavec_i}(\xivec)\Psi_{\alphavec_j}(\xivec)\right\rangle = 0$}{} with $\forall i\ne j$. Thus,
the PC coefficients are given by
\begin{equation*}
    c_{\alphavec_i} = \frac{\left\langle m(\mathcal{T}^{-1}(\xivec))\Psi_{\alphavec_i}(\xivec)\right\rangle}
                 {\left\langle\Psi_{\alphavec_i}^2(\xivec)\right\rangle}.
\end{equation*}
While the denominator is evaluated by analytic formulas (see examples in the appendix of \cite{SUDRET2008964}), the numerator is
computed by Gaussian quadrature integration where
\begin{equation*}
    N = (p+1)^d
\end{equation*}
integration points, i.e.\ high fidelity model $m(\xvec)$ evaluations, are required.

\subsection{Linear Regression Method}
The coefficients of \Eqref{eq:uq_approx} can also be computed with regression-based methods
\begin{equation}
    \boldsymbol{\hat{c}} = \argmin_{\cvec}\frac{1}{2}\norm{2}{\sum_{j=0}^{N-1}\left(m(\xvec_j) -
    \sum_{i=0}^{P-1}c_{\alphavec_i}\Psi_{\alphavec_i}(\xivec_j)\right)}^2
    + \frac{\lambda_{1}}{2}\norm{2}{\cvec}^2
    + \lambda_{2}\norm{1}{\cvec}
    \label{eq:uq_regression}
\end{equation}
\reviewchange{with regularization parameters $\lambda_{1},\lambda_{2}\ge 0$, and the $l_1$ norm and $l_2$ norm denoted by $\norm{1}{\cdot}$ and
$\norm{2}{\cdot}$, respectively}. The minimization problem is called ordinary least squares if
$\lambda_{1} = \lambda_{2} = 0$, elastic net \cite{doi:10.1111/j.1467-9868.2005.00503.x} if $\lambda_{1},\lambda_{2} > 0$, ridge regression \cite{doi:10.1080/00401706.1970.10488634} (or Tikhonov regularization) if only $\lambda_{1} > 0$ and Lasso \cite{doi:10.1111/j.2517-6161.1996.tb02080.x} if only $\lambda_{2} > 0$.
In matrix form the problem reads
\begin{equation}
    \boldsymbol{\hat{c}} = \argmin_{\cvec}\frac{1}{2}\norm{2}{\yvec - A\cvec}^2
    + \frac{\lambda_{1}}{2}\norm{2}{\cvec}^2
    + \lambda_{2}\norm{1}{\cvec}
    \label{eq:uq_regression_matrix_form}
\end{equation}
with model response $\yvec = \left(m(\xvec_0),\dots,m(\xvec_{N-1})\right)^\intercal\in\mathbb{R}^{N\times 1}$,
unknown coefficient vector $\cvec = \left(c_{\alphavec_0},\dots,c_{\alphavec_p}\right)^\intercal\in\mathbb{R}^{P\times 1}$
and system matrix $A\in\mathbb{R}^{N\times P}$. In case of $\lambda_{2} = 0$, the coefficients of \Eqref{eq:uq_regression_matrix_form}
are obtained in closed form by
\begin{equation*}
    \boldsymbol{\hat{c}} = \left(A^{\intercal}A + \lambda_{1} I\right)^{-1}A^{\intercal}\yvec
\end{equation*}
with $P\times P$ identity matrix $I$.
In contrast to the projection method (cf. \Secref{sec:uq_projection_method}), this method does
not require a fixed number of samples $N$. However, \cite{SUDRET2008964} gives an empirical optimal training sample size of
\begin{equation}
    N = (d-1)P
    \label{eq:uq_num_samples_regression}
\end{equation}
with $P$ defined in \Eqref{eq:num_TO_indices}.

\subsection{Orthogonal Matching Pursuit}
\label{sec:OMP}
The matching pursuit (MP) is a greedy algorithm developed by \cite{258082} which was enhanced by \cite{342465} to obtain
better convergence. This improved method is called orthogonal MP (OMP). In terms of PCE, the algorithm searches a minimal set
of non-zero coefficients to represent the model response, i.e.
\begin{equation*}
    \begin{aligned}
    \boldsymbol{\hat{c}} = \argmin_{\cvec} \quad & \norm{2}{\yvec - A\cvec}^2 \\
    \textrm{subject to} \quad & \norm{0}{\cvec} \le N_{c} \\
    \end{aligned}
\end{equation*}
\reviewchange{where $\norm{0}{\cvec}$ denotes the number of non-zero coefficients in $\cvec$ with a user-defined maximum $N_{c}$
\cite{OMP_Rubinstein}}. The vectors and matrices are defined
according to
\Eqref{eq:uq_regression_matrix_form}. \reviewchange{It is an iterative procedure where in the $(i+1)$-th
step a new coefficient vector $\cvec_{i+1}$ is searched that maximizes the inner product to the current residual
$\rvec_i = \yvec - \yvec_{i}$. We refer to the given references for details.}

\subsection{Bayesian Compressive Sensing}
\label{sec:BCS}
As stated in \cite{BCS_4524050, BCS_5256324}, the linear regression model \Eqref{eq:uq_regression} can be interpreted in a Bayesian
manner, i.e.
\begin{equation*}
    p\left(\cvec|\mathcal{D}\right) = \frac{p\left(\mathcal{D}\vert\cvec\right)p\left(\cvec\right)}{p\left(\mathcal{D}\right)}
\end{equation*}
with posterior distribution $p\left(\cvec|\mathcal{D}\right)$, likelihood $p\left(\mathcal{D}\vert\cvec\right)$, prior
$p\left(\cvec\right)$ and evidence $p\left(\mathcal{D}\right)$ of training data $\mathcal{D} = \{\xvec, y\}_{j=0}^{N-1}$
\cite{Sargsyan2017}. The likelihood
is assigned a Gaussian noise model
\begin{equation*}
    p\left(\mathcal{D}\vert\cvec\right) = \frac{1}{\left(2\pi\sigma^2\right)^{N/2}}
        \exp\left(-\sum_{j=0}^{N-1}\frac{\left(m\left(\xvec_i\right) - \hat{m}\left(\xvec_i\right)\right)^2}{2\sigma^2}\right)
\end{equation*}
with variance $\sigma^2$. It is a measure of how well the high fidelity model is represented by the
surrogate model \Eqref{eq:uq_approx}. In order to favour a sparse PCE solution, a Laplace prior
\begin{equation}
    p\left(\cvec\right) = \left(\frac{\lambda}{2}\right)^{P+1}
        \exp\left(-\lambda\sum_{i=0}^{P}\left|c_{i}\right|\right)
    \label{eq:uq_laplace_priors}
\end{equation}
is chosen. Using \Eqref{eq:uq_laplace_priors} in the maximum a posteriori (MAP) estimate for $\cvec$, i.e.
\begin{equation}
\argmax_{\cvec}{}\log\left[p\left(\mathcal{D}\vert\cvec\right)p\left(\cvec\right)\right],
    \label{eq:MAP}
\end{equation}
the Bayesian approach is
equivalent to \Eqref{eq:uq_regression} with $\lambda_{1} = 0$ \cite{NIPS2001_1976}, since \Eqref{eq:MAP} is identical to
a minimization of
\begin{equation*}
    \argmin_{\cvec}{}-\log\left[p\left(\mathcal{D}\vert\cvec\right)p\left(\cvec\right)\right].
\end{equation*}
An iterative algorithm to obtain the coefficients is described in
\cite{BCS_5256324}. It requires a user-defined stopping threshold $\varepsilon$ that basically controls the number of kept basis
terms, with more being skipped the higher the value is. The overall method is known as Bayesian Compressive Sensing (BCS)
\cite{BCS_4524050, BCS_5256324}.

\subsection{Sensitivity Analysis}
\label{sec:Sensitivity_Analysis}
Sobol' indices \cite{SOBOL2001271} are good measures of sensitivity since they provide information about single and mixed parameter
effects. In addition to these sensitivity measures, there are various other methodologies such as
Morris screening
\cite{doi:10.1080/00401706.1991.10484804}. A survey is presented in \cite{GAN2014269} on the example of a hydrological model.
Instead of Monte Carlo, Sobol' indices are also easily obtained by surrogate models based on PCE as discussed in the
following subsections.

\subsubsection{Sobol' Sensitivity Indices}
In \cite{SOBOL2001271} Sobol' proposed global sensitivity indices that are calculated on an analysis of
variance (ANOVA) decomposition (Sobol' decomposition) of a square integrable function $f(\xvec)$ with
$\xvec\in I^d \coloneqq [0, 1]^d$, i.e. \cite{Sobol1990} 
\begin{equation}
    f(\xvec) = f_0
             + \sum_{i=1}^{d}f_{i}(x_i)
             + \sum_{1\le i_1 < \dots < i_s \le d}f_{i_{1}i_{s}}(x_{i_{1}}, x_{i_{s}})
             + \dots
             + f_{12\dots d}(x_1, x_2, \dots, x_d)
    \label{uq:sobol_decomposition}
\end{equation}
with mean
\begin{equation*}
    f_0 = \int_{I^d}f(\xvec)\dx{\xvec}
\end{equation*}
and
\begin{equation}
    \int_{0}^{1}f_{i_1\dots i_s}(x_{i_1},\dots, x_{i_s})\dx{x_k} = 0,
    \label{eq:uq_sobol_integral}
\end{equation}
for $k= i_1,\dots,i_s$ and $s=1,\dots,d$.
Since \Eqref{eq:uq_sobol_integral} holds, the components of \Eqref{uq:sobol_decomposition} are mutually orthogonal. Therefore,
the total variance of \Eqref{uq:sobol_decomposition} is
\begin{equation*}
    D = \int_{I^d}f^2(\xvec)\dx{\xvec} - f_0^2
\end{equation*}
that can also be written as
\begin{equation}
    D = \sum_{i=1}^{d}D_{i} + \sum_{1\le i_1 < \dots < i_s \le d}D_{i_{1}i_{s}} + \dots + D_{123\dots d}
    \label{eq:uq_sobol_total_variance_split}
\end{equation}
where
\begin{equation}
    D_{i_1\dots i_s} = \int_{I^s}f_{i_1\dots i_s}^{2}(x_{i_1},\dots, x_{i_s})\dx{x_{i_1}}\cdots\dx{x_{i_s}},
    \label{eq:uq_sobol_specific_variance}
\end{equation}
with $1 \le i_1 < \dots < i_s \le d$. Based on \Eqref{eq:uq_sobol_total_variance_split} and \Eqref{eq:uq_sobol_specific_variance}, the
Sobol' indices are defined as
\begin{equation*}
    S_{i_1\dots i_s} \coloneqq \frac{D_{i_1\dots i_s}}{D}
\end{equation*}
with
\begin{equation}
    \sum_{i=1}^{d}S_{i} + \sum_{1\le i < j\le d} S_{ij}+\dots+S_{12\dots d} = 1.
    \label{eq:sobol_sum}
\end{equation}
The first order indices $S_i$ are also known as main sensitivities. They describe the effect of a single input parameter on the
model response. The total effect of the $i$-th design variable on the model response, proposed by \cite{HOMMA19961}, is the sum
of all Sobol' indices that include the $i$-th index, i.e.
\begin{equation*}
    S_{i}^{T} = \sum_{\boldsymbol{i}\in\mathcal{I}}S_{\boldsymbol{i}}
\end{equation*}
with $\mathcal{I} = \{\boldsymbol{i} = (i_1,\dots, i_s): \exists k, 1 \le k \le s \le d, i_k = i\}$.

\subsubsection{Sobol' Indices using Polynomial Chaos Expansion}
Instead of Monte Carlo techniques, Sobol' indices can be estimated using surrogate models based on PCE since the truncated
expansion can be rearranged like
\Eqref{uq:sobol_decomposition}. The Sobol' estimates are then given by \cite{SUDRET2008964}
\begin{equation*}
    \hat{S}_{i_1\dots i_s} = \frac{1}{\hat{D}}\sum_{\alphavec\in\mathcal{I}_{i_1,\dots, i_s}}
                                c_{\alphavec}^{2}\left\langle\Psi_{\alphavec}^2\right\rangle
\end{equation*}
where
\begin{equation*}
    \mathcal{I}_{i_1,\dots,i_s} = \left\{\alphavec:
    \begin{matrix}
        \alpha_k > 0 & \forall k = 1,\dots, n, \quad k\in\left(i_1,\dots, i_s\right) \\
        \alpha_k = 0 & \forall k = 1,\dots, n, \quad k\not\in\left(i_1,\dots, i_s\right)
    \end{matrix}
    \right\}
\end{equation*}
and variance
\begin{equation*}
    \hat{D} = \sum_{i=1}^{P-1}c_{\alphavec_i}^{2}\left\langle\Psi_{\alphavec_i}^{2}\right\rangle.
\end{equation*}
The main and total sensitivities are computed by
\reviewchange{
\begin{equation*}
\hat{S}_{i} = \frac{1}{\hat{D}}  \sum_{\alphavec\in\mathcal{I}_{i}}c_{\alphavec}^2\moment{\Psi_{\alphavec}^2}
\end{equation*}
with $\mathcal{I}_{i} = \{\alphavec=\left(\alpha_1,\dots,\alpha_d\right) : \alpha_i > 0 \wedge \forall k \ne i, \alpha_k = 0\}$}{}
and
\begin{equation*}
\hat{S}_{i}^{T} = \frac{1}{\hat{D}} \sum_{\alphavec\in\mathcal{I}_i} c_{\alphavec}^2\moment{\Psi_{\alphavec}^2},
\end{equation*}
with \reviewchange{$\mathcal{I}_{i} = \{\alphavec = (\alpha_1,\dots,\alpha_d) : \alpha_i > 0\}$}, respectively.

\subsection{Confidence Intervals using Bootstrap}
\label{sec:ci_bootstrap}
In this subsection we briefly outline the computation of confidence intervals for the estimates of Sobol'
indices using the bootstrap method \cite{efron1979}. In the context of PCE, the bootstrap method has already been applied in
\cite{MARELLI201867}, where it is referred to as bootstrap-PCE (or bPCE).  \reviewchange{The bootstrap method, in general, generates $B$ independent
samples each of size $N$ by resampling from the original dataset. Each bootstrap sample, that may contain a point several times, is
then considered as a new training sample to compute the coefficients of \Eqref{eq:uq_approx}}.
In order to calculate the \SI{95}{\percent} confidence interval for Sobol' indices we follow the description of \cite{doi:10.1080/00949659708811825},
where \reviewchange{the bounds are given by}
\begin{equation*}
    \hat{S}_{i_1\dots i_s} \pm 1.96 \cdot s.e.(\hat{S}_{i_1\dots i_s})
\end{equation*}
with $1\le s\le d$ and standard error ($s.e.$) of $B\in\mathbb{N}^{>1}$ bootstrap samples
\begin{equation*}
    s.e.(\hat{S}_{i_1\dots i_s}) = \sqrt{\frac{1}{B-1}\sum_{b=1}^{B}\left(S_{i_1\dots i_s}^{(b)} - S_{i_1\dots i_s}^{*}\right)^2}
\end{equation*}
and bootstrap sample mean
\begin{equation*}
    S_{i_1\dots i_s}^{*} = \frac{1}{B}\sum_{b=1}^{B}S_{i_1\dots i_s}^{(b)}.
\end{equation*}

\section{Experiment Design}
\label{sec:experiment_design}

\subsection{High Intensity Cyclotrons}
In order to study the effect of AMR solver parameters and energy binning in neighbouring bunch simulations we perform sensitivity
experiments with three different high intensity cyclotrons, the \ringcyc{} \cite{PhysRevAccelBeams.22.064602}, the
\injII{} \cite{KOLANO201854} and the \daedalus{} Injector Cyclotron
(DIC) \cite{doi:10.1063/1.4802375, doi:10.1063/1.4826879}. We always accelerate 5 particle bunches with
$10^6$ particles.
The coarsest level grid is kept constant with $24^3$ mesh points which is refined twice. For the \injII{} and \ringcyc{}
the particles are integrated in time over one turn using \num{2880} steps per turn and for the DIC over three turns with \num{1440}
steps per turn. The experimental setup is summarized in \Tabref{tab:experimental_setup}.
In all experiments the initial particle distribution is read from a checkpoint file to guarantee identical conditions for all
training and validation points of a UQ sample.

\begin{table}[!ht]
    \centering
    \sisetup{scientific-notation=true}
    \begin{tabular}{cccccc}
        \toprule
        no. turns   & steps/turn    & no. bunches   & particles/bunch   & PIC base grid & no. AMR levels \\
        \midrule
        1 or 3      & 1440 or 2880  & 5             & $10^{6}$          & $24\times 24\times 24$ & 2 \\
        \bottomrule
    \end{tabular}
    \caption{Experimental setup of the \ringcyc{}, \injII{} and \daedalus{} Injector Cyclotron model.}
    \label{tab:experimental_setup}
\end{table}

A list of the design variables under consideration is given in \Tabref{tab:dvar_sampling_injII}. While the resolution is basically
controlled by the maximum number of AMR levels, the refinement policy affects its location. As described in \cite{FREY2019106912}
the \opal{} library provides several refinement criteria such as the charge density per grid point, the potential as well as the
electric field. Here, we want to analyze the effect of the threshold $\lambda\in[0, 1]$ of the electrostatic
potential refinement policy, where a grid cell $(i,j,k)$ on a level $l$ is refined if
\reviewchange{
\begin{equation*}
    |\phi^{l}_{i,j,k}|\ge \lambda \max_{i,j,k} |\phi^{l}_{i,j,k}|
\end{equation*}}
holds. Due to the motion of the particles in space the multi-level hierarchy has to be updated regularly to maintain the resolution
which is defined by the regrid frequency $f_{r}$. It should be noted that the regrid frequency defines the number of steps until
the AMR hierarchy is updated. Hence, if $f_{r} = 1$, the
AMR levels are updated in each time step. Whenever this happens, the electric self-field needs to be
recalculated by solving Poisson's equation.
The number of Poisson solves 
is controlled
by the number of energy bins and therefore by the binning parameter $\eta$ (cf. \Secref{sec:numerical_model}).
The lower the value of $\eta$, the smaller the bin width and, hence, the more expensive the model is.

\begin{figure}[!ht]
    \includegraphics[width=1.0\textwidth]{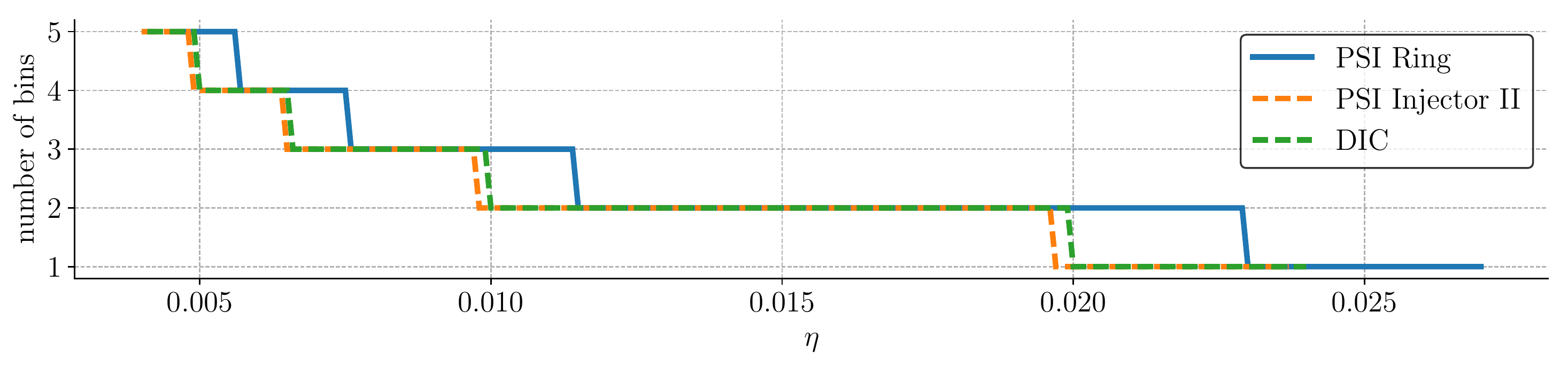}
    \caption{Number of energy bins in the \ringcyc, \injII{} and \daedalus{} Injector Cyclotron (DIC) with respect to the binning
    parameter $\eta$. The shown binning curves are with respect to the initial simulation energies used in this study.
    A straightforward uniform sampling from the full range yields a biased sample set.}
    \label{fig:binning}
\end{figure}

As an upper limit of the regrid frequency $f_r$ we choose \num{120} integration steps. Since we perform either
\num{1440} or \num{2880} steps per turn, this corresponds to an azimuthal angle of
\SI{30}{\degree} and \SI{15}{\degree}, respectively.
The choice of the binning parameter $\eta$ in \Eqref{eq:binning} depends mainly on the energy
difference between bunches. The upper bound of the sampling range was selected such that we have at most
as many energy bins as bunches in simulation. However, the sampling from the range is not straightforward since there exist
more states with fewer energy bins (cf. \Figref{fig:binning}) due to \Eqref{eq:binning}. Instead, we sample the binning
parameter in subranges of equal bin count in order to avoid a biased sample set.

\begin{table}[!ht]
    \centering
    \sisetup{scientific-notation=true}
    \begin{tabular}{lll}
        \toprule
        symbol                  & design variable           & sampling range \\
        \midrule
        $f_{r}$                 & regrid frequency          & $[1, 120]$ \\
        $\lambda$               & refinement threshold      & $[0.5, 0.9]$ \\
        \multirow{3}{*}{$\eta$} & binning Ring $(10^{-3})$   & $[4.7, 5.7]\cup
                                                               [5.8, 7.6]\cup
                                                               [7.7, 11.5]\cup
                                                               [11.6, 23.0]\cup
                                                               [23.1, 27.1]$ \\
                                & binning Inj-2 $(10^{-3})$ & $[4.0, 4.9]\cup
                                                               [5.0, 6.5]\cup
                                                               [6.6, 9.8]\cup
                                                               [9.9, 19.7]\cup
                                                               [19.8, 23.8]$ \\
                                & binning DIC $(10^{-3})$   & $[4.1, 5.0]\cup
                                                               [5.1, 6.6]\cup
                                                               [6.7, 10.0]\cup
                                                               [10.1, 20.0]\cup
                                                               [20.1, 24.1]$ \\
        \bottomrule
    \end{tabular}
    \caption{List of design variables and their sampling ranges for the neighbouring bunch simulations.} 
    \label{tab:dvar_sampling_injII}
\end{table}

\subsection{RF Electron Gun Model}
Like in the neighbouring bunch model, the time-to-solution in the rf electron gun model is dominated by the Poisson solver and
the time integration. A reduction of the computational effort with regard to the Poisson solver
is achieved by smaller PIC meshes and fewer energy bins $N_{E}$. The costs of the time integrator is cheapened with
coarser time steps $\Delta t$ and fewer macro particles. Instead of the AMR model, the rf electron gun model uses the
Fast Fourier Transform (FFT) Poisson solver of \opal{} where we put a $L_{x}\times L_{y}\times L_{z}$ uniform mesh
of $L_{x} = L_{y} = 64$ and $L_{z} = 32$ grid points. The final number of emitted macro particles is given by
\begin{equation*}
    N_{p} = p_{f}L_{x}L_{y}L_{z}
\end{equation*}
where the particle multiplication factor $p_{f}$ is an integer. This parameter basically controls the number
of particles per grid cell and, hence, the noise of the PIC model.
The design variables and sampling ranges are given in \Tabref{tab:dvar_sampling_awa}.
We model the rf electron gun of the Argonne Wakefield Accelerator (AWA) that has a length of approximately \SI{30}{cm}.
\begin{table}[!ht]
    \centering
    \begin{tabular}{lll}
        \toprule
        symbol      & design variable               & sampling range \\
        \midrule
        $p_{f}$     & particle factor               & $[1, 5]$ \\
        $N_{E}$     & number of bins                & $[2, 10]$ \\
        $\Delta t$  & time step $(\SI{0.1}{ps})$    & $[1, 10]$ \\
        \bottomrule
    \end{tabular}
    \caption{List of design variables and their sampling ranges for the rf electron gun model. The time step is the only
    floating point variable.}
    \label{tab:dvar_sampling_awa}
\end{table}

\subsection{Surrogate Model Selection}
In order to avoid overfitting we proceed like \cite{Sargsyan_2014} where the truncation order of
the PC expansion and the settings of the regression models are chosen such that the relative $l_2$ error
\reviewchange{
\begin{equation}
    \sqrt{\frac{\sum_{i=0}^{N-1}\left[m(\xvec_i) - \hat{m}(\xvec_i)\right]^2}{\sum_{i=0}^{N-1}m^2(\xvec_i)}}
    \label{eq:rel_l2_error}
\end{equation}}
between the surrogate
$\hat{m}(\xvec)$ and high fidelity model $m(\xvec)$ of the training and validation set is approximately of equal magnitude.
As an additional error measure we also compare the relative $l_1$ error
\reviewchange{
\begin{equation}
    \frac{\sum_{i=0}^{N-1}\left|m(\xvec_i) - \hat{m}(\xvec_i)\right|}
    {\sum_{i=0}^{N-1}\left|m(\xvec_i)\right|}.
    \label{eq:rel_l1_error}
\end{equation}}
The number of samples $N$ in \Eqref{eq:rel_l2_error} and \Eqref{eq:rel_l1_error}
corresponds either to the number of training $N_t$ or validation points $N_v$.
The total number of $N = 100$ samples was randomly partitioned into disjoint training and validation sets
with $N_t = 0.5N$ and $N_v = 0.5N$, respectively. Since we have $d=3$ design variables, we satisfy
\Eqref{eq:uq_num_samples_regression} with $N_{t}=50$ up to polynomial order $p=3$.



\section{Results}
\label{sec:results}
The estimated sensitivities are obtained from PC surrogate models where we use
either ordinary least squares (OLS) and Bayesian compressive sensing (BCS) of \uqtk{}
\cite{doi:10.1137/S1064827503427741, Debusschere2017} or orthogonal matching pursuit (OMP) of scikit-learn
\cite{scikit-learn, sklearn_api} together with Chaospy \cite{FEINBERG201546} to compute the expansion coefficients. A summary
of the PC model setups is given in \Tabref{tab:surrogate_model_summary}. In order to study the evolution of the sensitivities
we construct the PC surrogate models at equidistant steps of the accelerator models and
evaluate their sensitivities. These steps correspond to azimuthal angles in the cyclotrons
or longitudinal positions in the rf electron gun model.
In the examples below we only show the first order Sobol' indices since their sum is already almost one which is the maximum
per definition (cf. \Eqref{eq:sobol_sum}).






\begin{table}[!ht]
    \centering
    \sisetup{
             table-format = 1.0e-1,
             table-text-alignment = center,
             round-mode=places,
             round-precision=0,
             scientific-notation=true,
             fixed-exponent=0}
    \begin{tabular}{lcSc}
    \toprule
    model   & PC order & \multicolumn{2}{c}{stopping criterion} \\ \cmidrule(ll){3-4}
            &          & {$\varepsilon$ (BCS)} & $N_{c}$ (OMP) \\
    \midrule
    \ringcyc & 2 & 1.0e-7 & 5 \\ \midrule
    \injII   & 2 & 1.0e-4 & 7 \\ \midrule
    DIC      & 2 & 1.0e-8 & 6 \\ \midrule
    AWA      & 2 & 1.0e-9 & 7 \\ \bottomrule
    \end{tabular}
    \caption{PC surrogate model settings for all accelerator model examples. The stopping criterion of
    Bayesian Compressive Sensing (BCS) and Orthogonal Matching Pursuit (OMP) is discussed in
    \Secref{sec:BCS} and \Secref{sec:OMP}, respectively.}
    \label{tab:surrogate_model_summary}
\end{table}

\subsection{High Intensity Cyclotrons}
\begin{figure}[!ht]
    \includegraphics[width=1.0\textwidth]{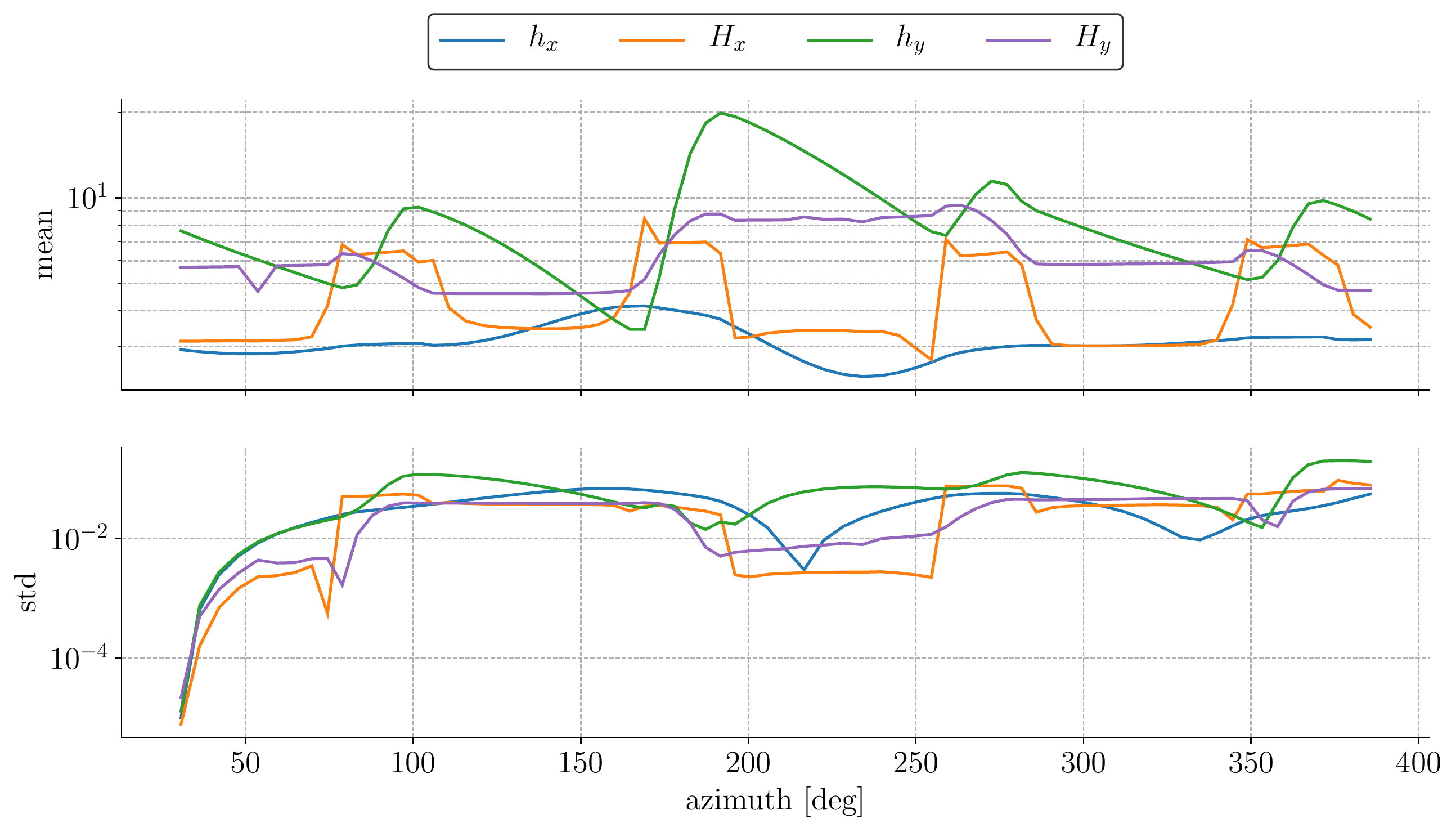}
    \caption{Evolution of the mean and standard deviation (std) \reviewchange{of the spatial-profile parameters $h_x$, $h_y$ and the phase-space
    halo parameters $H_x$, $H_y$ as defined in \Eqref{eq:profile_parameter} and \Eqref{eq:halo_parameter}, respectively. Based on the
    mean of $H_x$, the location of the dipoles in the \injII{} can be detected, i.e. at \SI{90}{\degree}, \SI{180}{\degree},
    \SI{270}{\degree} and \SI{360}{\degree}.}}
    \label{fig:injector2_evolution_mean_std}
\end{figure}

\begin{figure}[!ht]
    \includegraphics[width=1.0\textwidth]{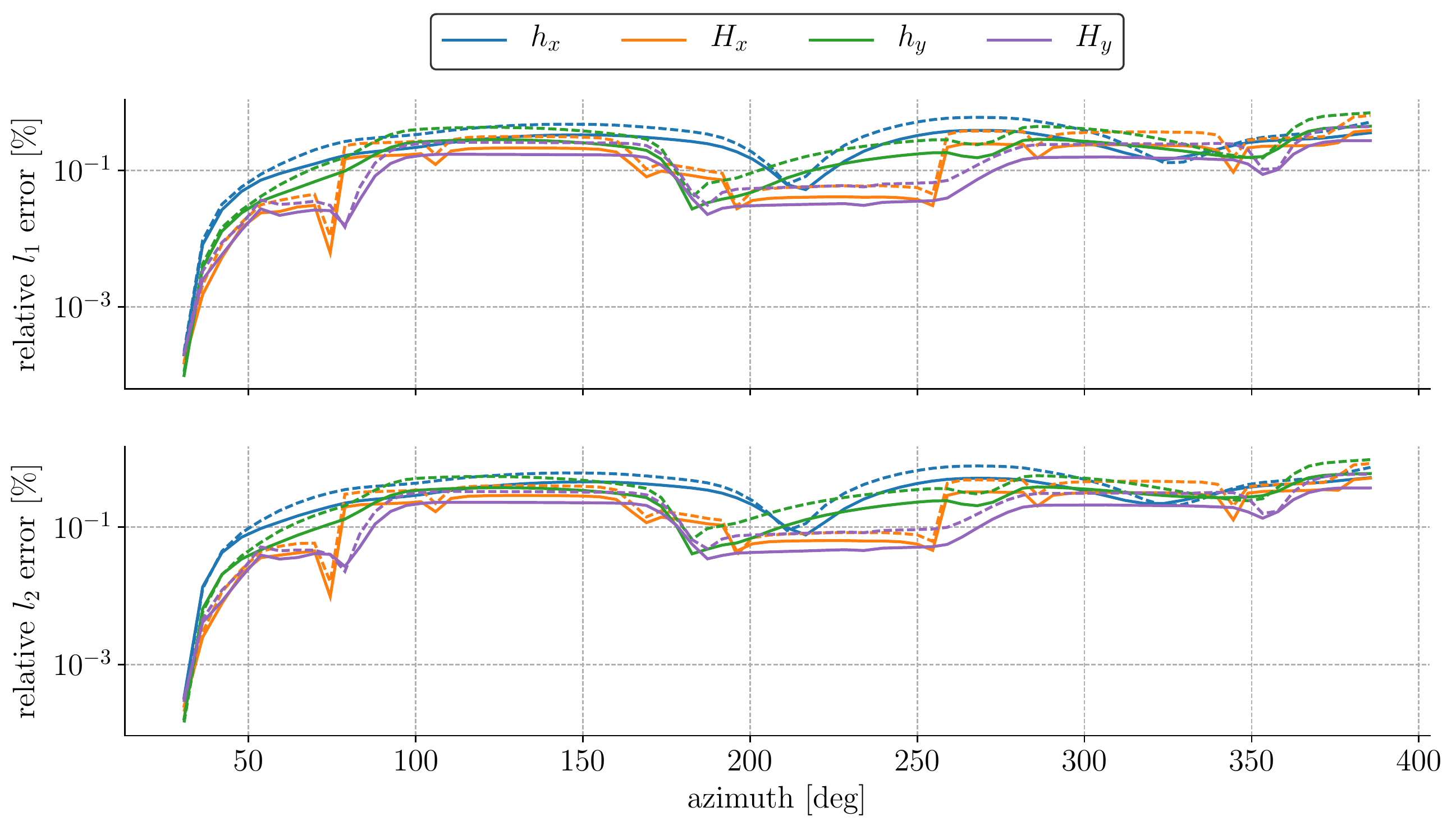}
    \caption{Evolution of the relative $l_2$ and $l_1$ error \reviewchange{between the surrogate and the true model}{} of the \injII.
    \reviewchange{The full lines are the errors to the surrogate model obtained with the training set and the dashed lines are the errors to the
    surrogate model computed with the validation set. For each quantity, the dashed and full lines are close to each other,
    indicating no overfitting of the surrogate model.}}
    \label{fig:injector2_evolution_training_validation_error}
\end{figure}

\begin{figure}[!ht]
    \centering
    \includegraphics[width=0.8\textwidth]{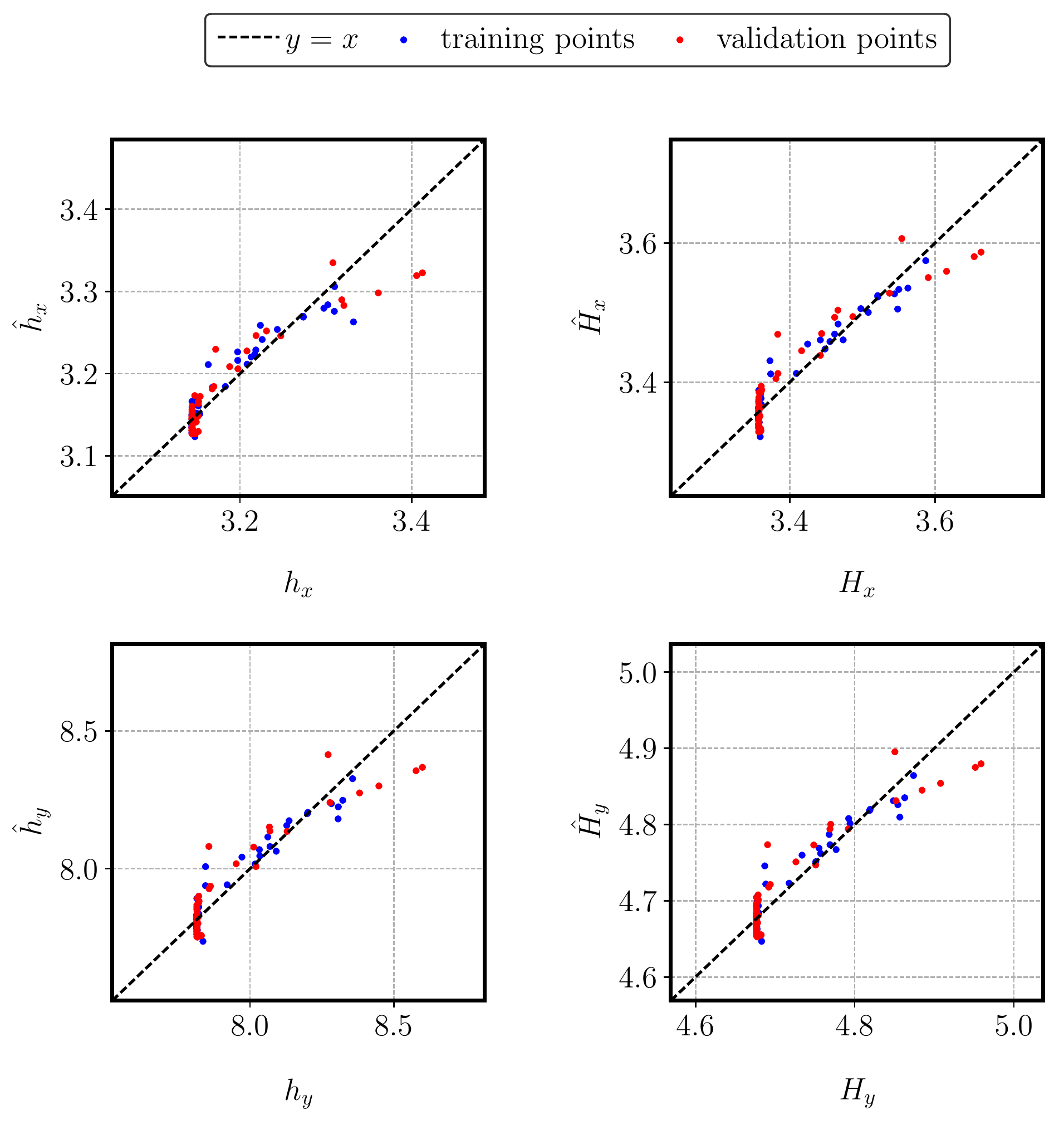}
    \caption{Comparison between the high fidelity ($x$-axis) and PC surrogate model ($y$-axis) at
    \SI[round-mode=places, round-precision=0]{389.76}{\degree} of the \injII{} simulation.
    The blue and red dots indicate the training and validation points, respectively. In the best case all points coincide with
    the dashed black line.}
    \label{fig:injector2_high_fidelity_vs_surrogate_step_391}
\end{figure}

\begin{figure}[!ht]
    \includegraphics[width=1.0\textwidth]{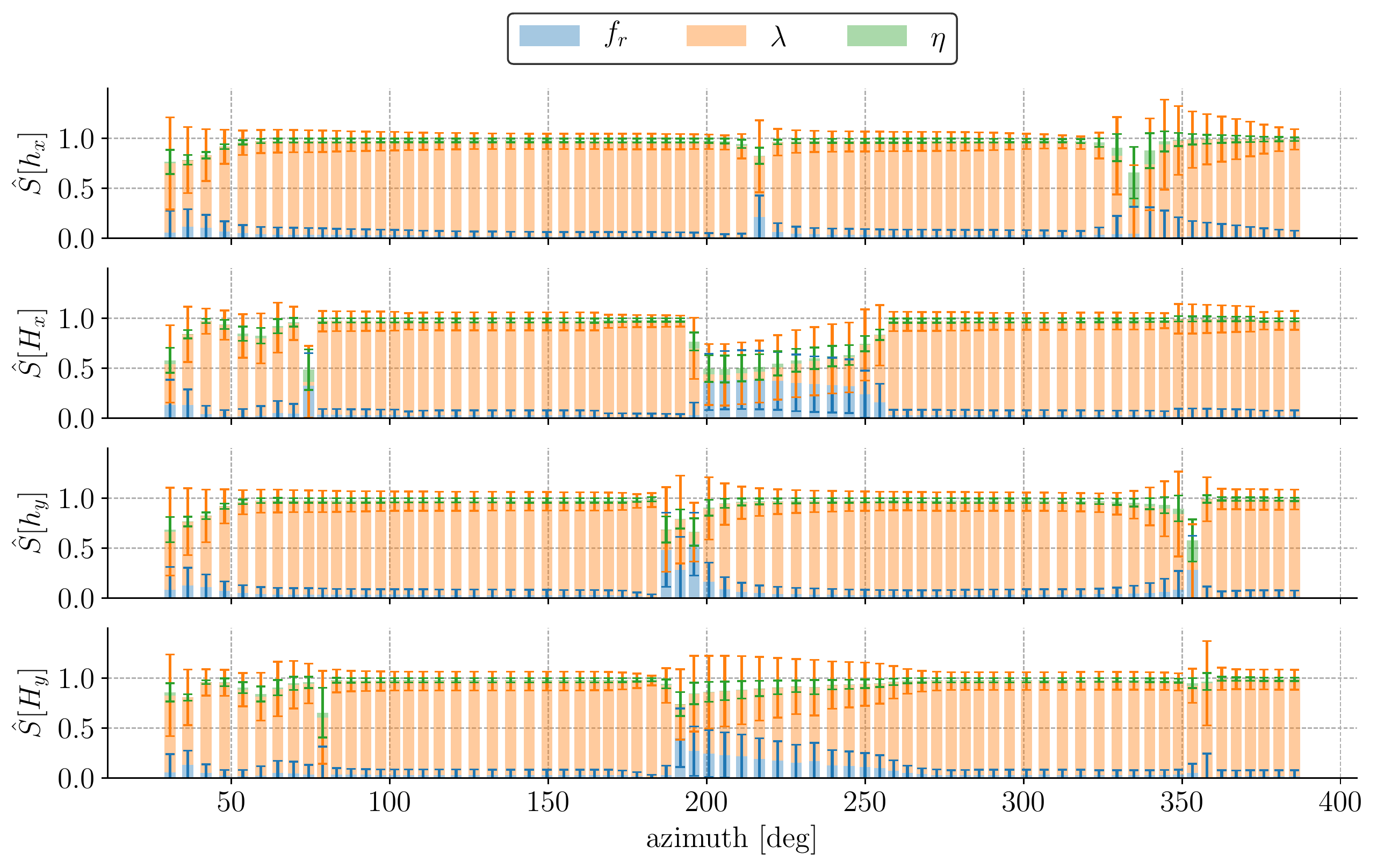}
    \caption{Evolution of the estimated first order Sobol' indices in the \injII.
    The error bars denote the \SI{95}{\percent} bootstrapped $(B=100)$ confidence interval (cf. \Secref{sec:ci_bootstrap}).
    \reviewchange{The main sensitivities are evaluated for the spatial-profile parameters $h_x$, $h_y$ and the phase-space halo parameters $H_x$,
    $H_y$ as defined in \Eqref{eq:profile_parameter} and \Eqref{eq:halo_parameter}, respectively. The bars are coloured with respect
    to the regrid frequency $f_r$, AMR refinement threshold $\lambda$ and energy binning parameter $\eta$. The
    refinement threshold has the highest impact on the halo measures.}}
    \label{fig:injector2_evolution_main_sensitivities}
\end{figure}

\begin{figure}[!ht]
    \includegraphics[width=1.0\textwidth]{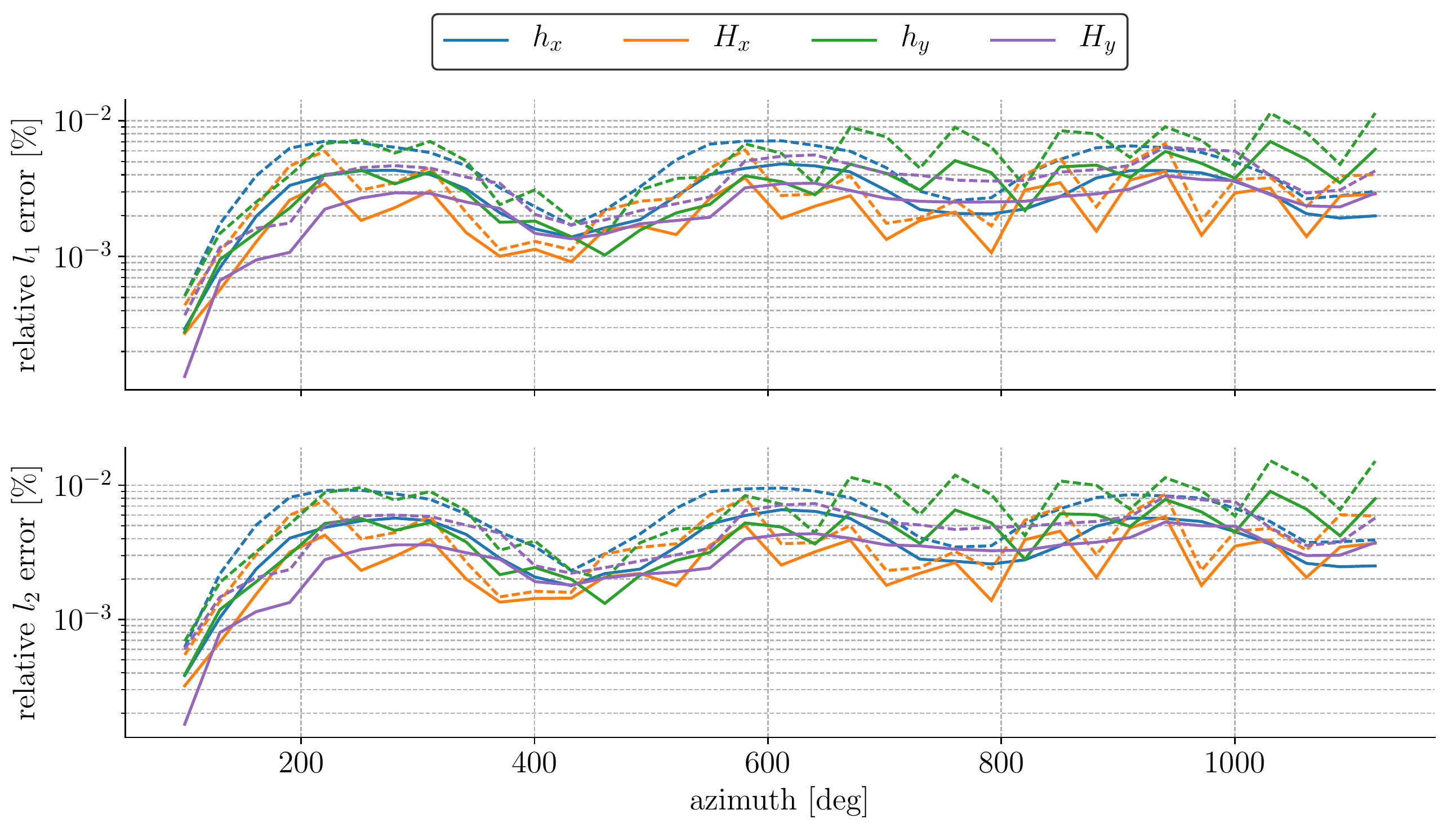}
    \caption{Evolution of the relative $l_2$ and $l_1$ error \reviewchange{between the surrogate and the true model}{} of the \daedalus{} Injector
    Cyclotron. \reviewchange{The full lines are the errors to the surrogate model obtained with the training set and the dashed lines
    are the errors to the surrogate model computed with the validation set. For each quantity, the dashed and full lines are close to
    each other, indicating no overfitting of the surrogate model.}}
    \label{fig:isodar_evolution_training_validation_error}
\end{figure}

\begin{figure}[!ht]
    \centering
    \includegraphics[width=0.8\textwidth]{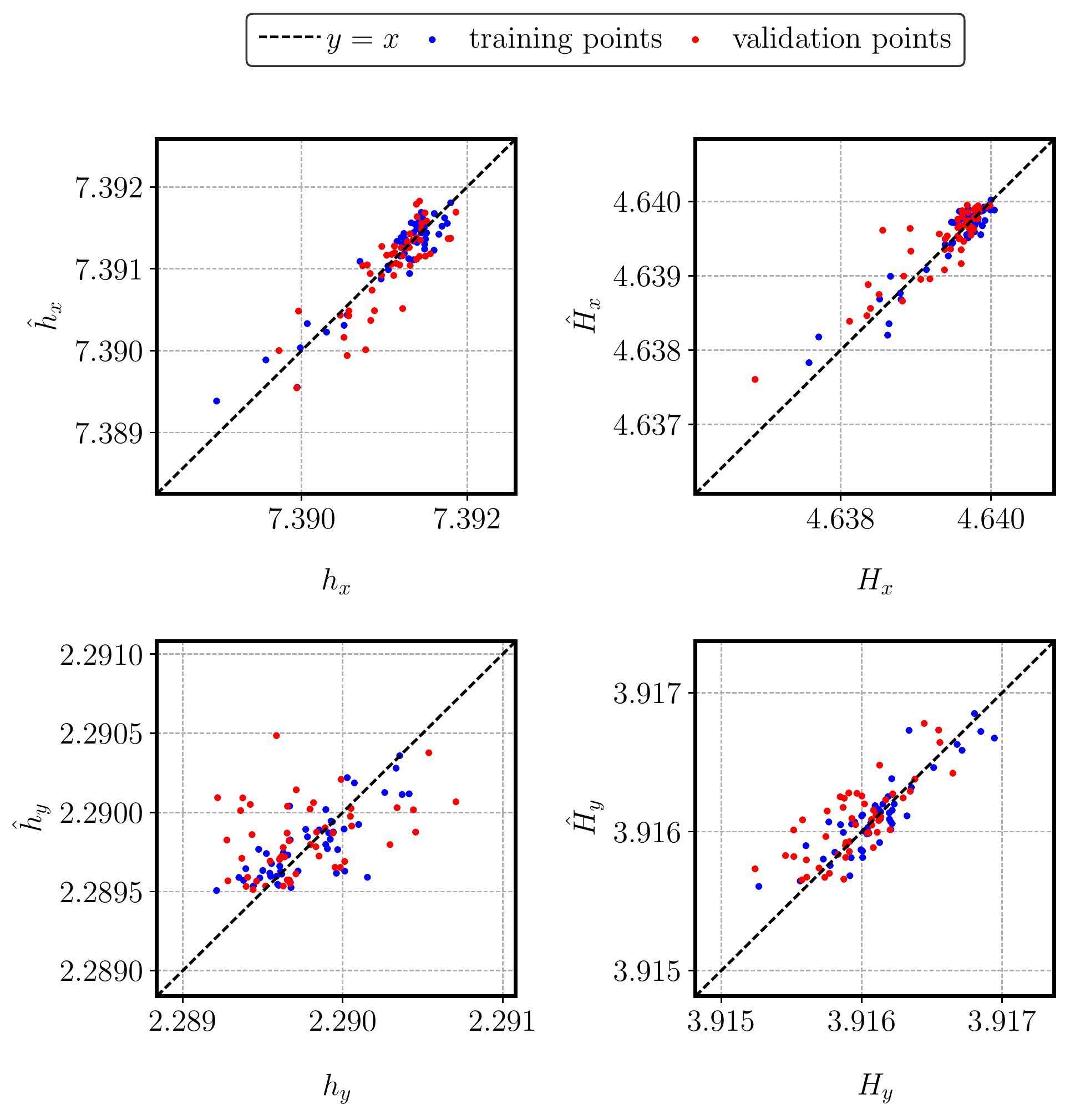}
    \caption{Comparison between the high fidelity ($x$-axis) and PC surrogate model ($y$-axis) at
    \SI[round-mode=places, round-precision=0]{1120.395}{\degree} of the \reviewchange{\daedalus{} Injector Cyclotron}{} simulation.
    The blue and red dots indicate the training and validation points, respectively. In the best case all points coincide with
    the dashed black line.}
    \label{fig:isodar_high_fidelity_vs_surrogate_step_107}
\end{figure}

\begin{figure}[!ht]
    \includegraphics[width=1.0\textwidth]{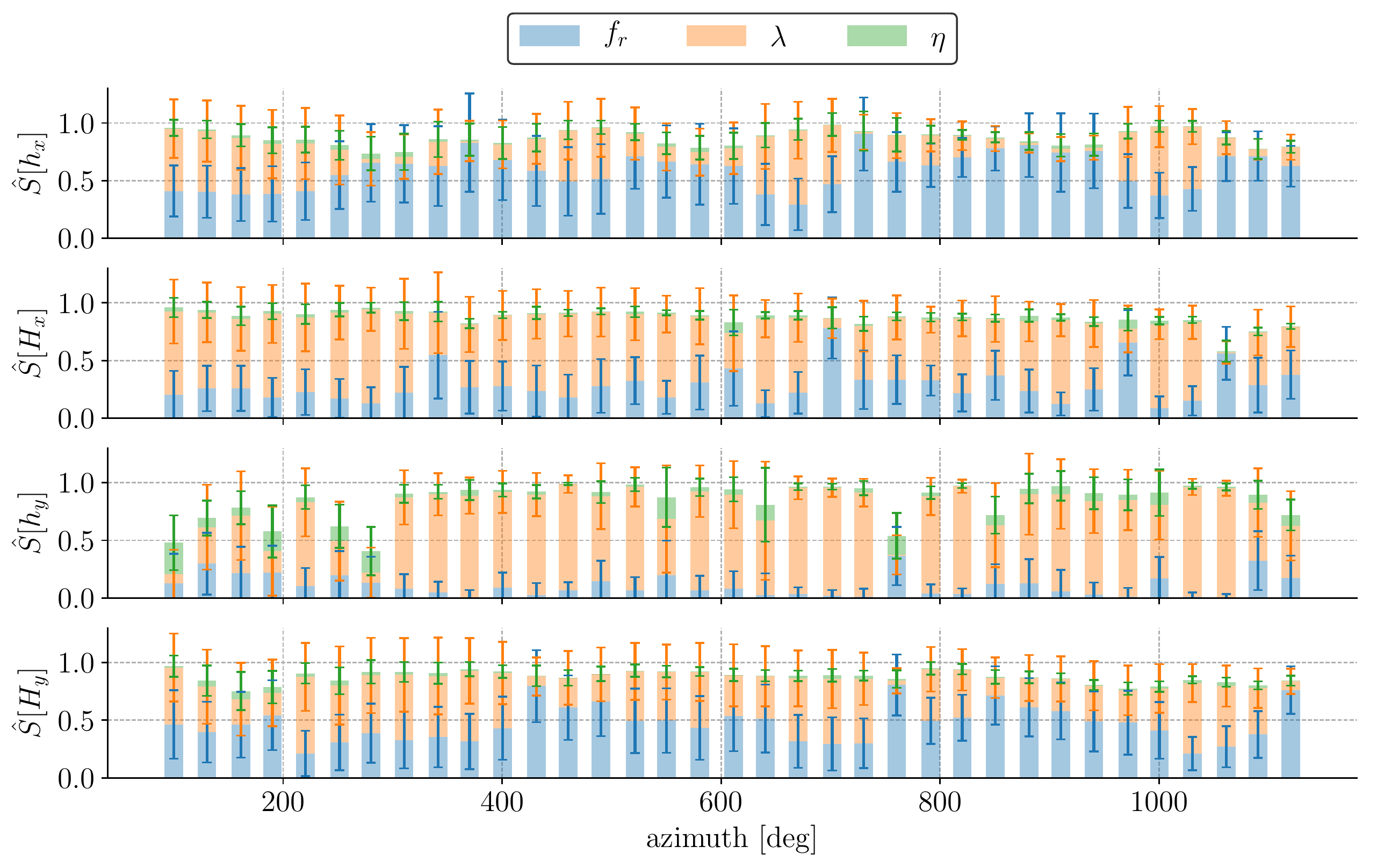}
    \caption{\reviewchange{Evolution of the estimated first order Sobol' indices in the \daedalus{} Injector Cyclotron. The error bars denote the
    \SI{95}{\percent} bootstrapped $(B=100)$ confidence interval (cf. \Secref{sec:ci_bootstrap}). The main sensitivities are
    evaluated for the spatial-profile parameters $h_x$, $h_y$ and the phase-space halo parameters $H_x$, $H_y$ as defined
    in \Eqref{eq:profile_parameter} and \Eqref{eq:halo_parameter}, respectively. The bars are coloured with respect to the regrid
    frequency $f_r$, AMR refinement threshold $\lambda$ and energy binning parameter $\eta$. Beside the refinement threshold, the
    regrid frequency has also a significant impact on the halo measures.}}
    \label{fig:isodar_evolution_main_sensitivities}
\end{figure}

\begin{figure}[!ht]
    \includegraphics[width=1.0\textwidth]{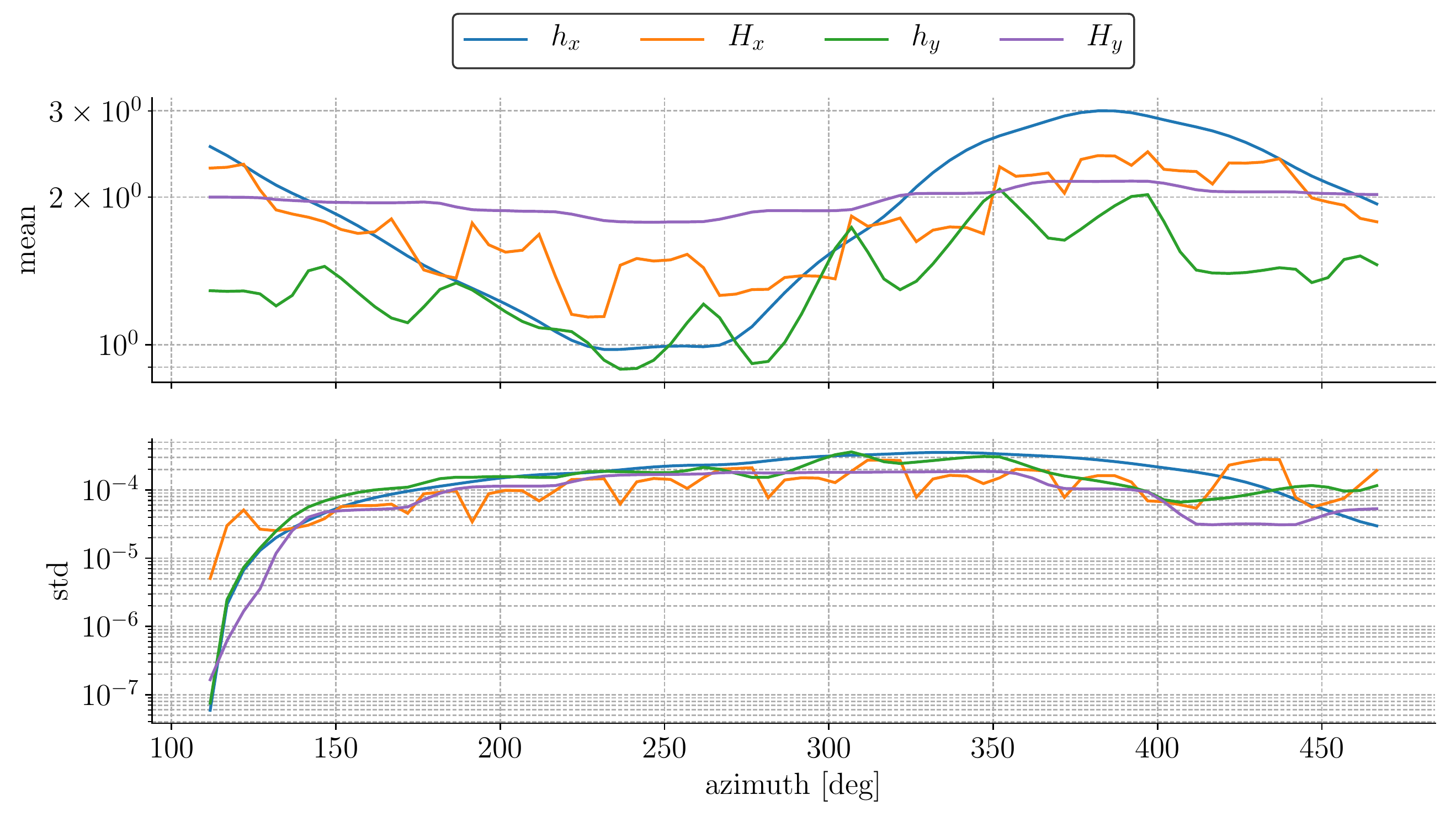}
    \caption{Evolution of the mean and standard deviation (std) \reviewchange{of the spatial-profile parameters $h_x$, $h_y$ and the phase-space
    halo parameters $H_x$, $H_y$ as defined in \Eqref{eq:profile_parameter} and \Eqref{eq:halo_parameter}, respectively. The
    variability of these quantities in the \ringcyc{} cyclotron is on the order of $\mathcal{O}(10^{-4})$ which is two orders of magnitude smaller than for the \injII{} (cf. \Figref{fig:injector2_evolution_mean_std}).}}
    \label{fig:ring_evolution_mean_std}
\end{figure}

\begin{figure}[!ht]
    \includegraphics[width=1.0\textwidth]{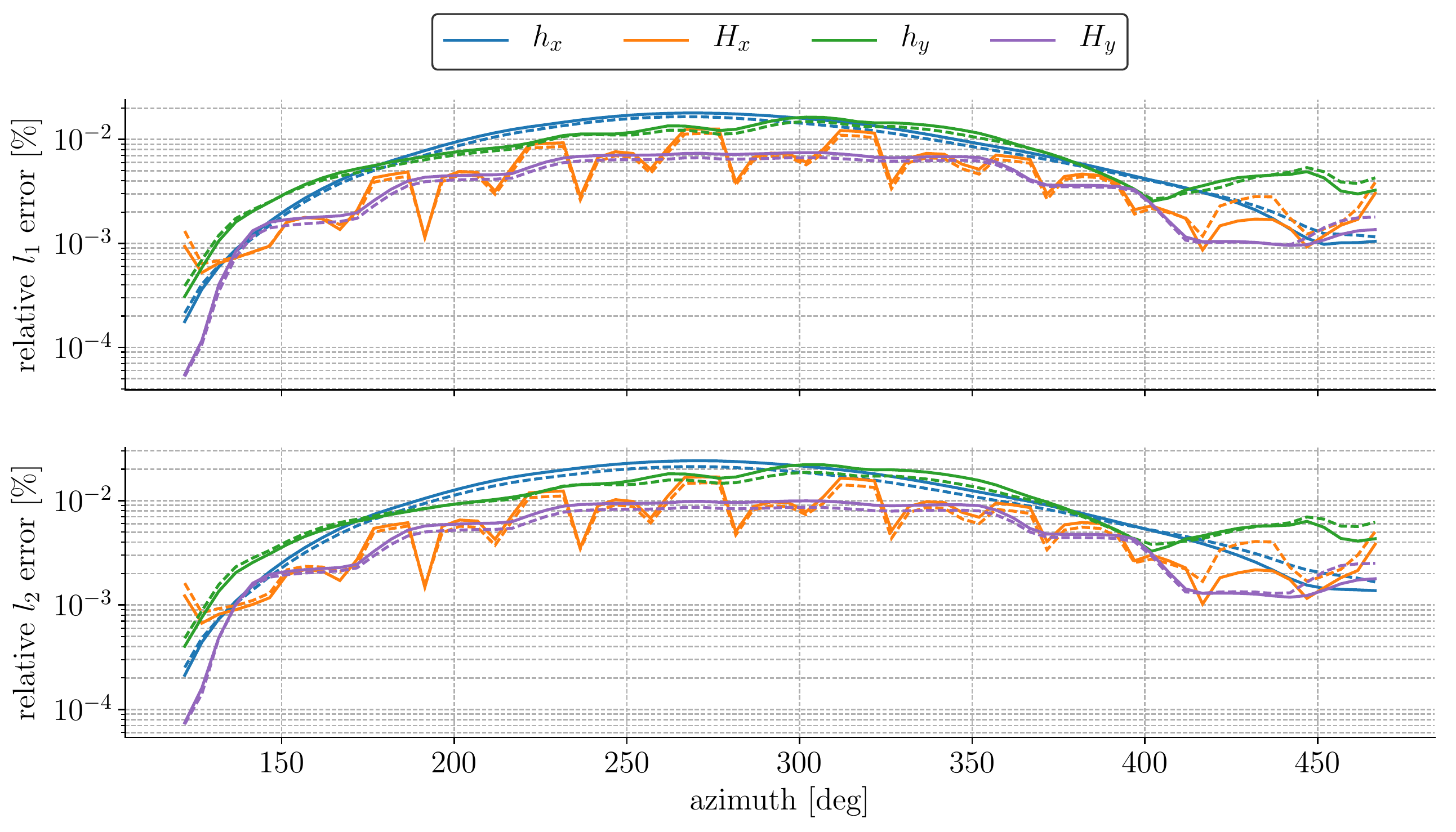}
    \caption{Evolution of the relative $l_2$ and $l_1$ error \reviewchange{between the surrogate and the true model}{} of the \ringcyc{} cyclotron.
    \reviewchange{The full lines are the errors to the surrogate model obtained with the training set and the dashed lines are the errors to the
    surrogate model computed with the validation set. For each quantity, the dashed and full lines are close to each other, indicating
    no overfitting of the surrogate model.}}
    \label{fig:ring_evolution_training_validation_error}
\end{figure}

\begin{figure}[!ht]
    \centering
    \includegraphics[width=0.8\textwidth]{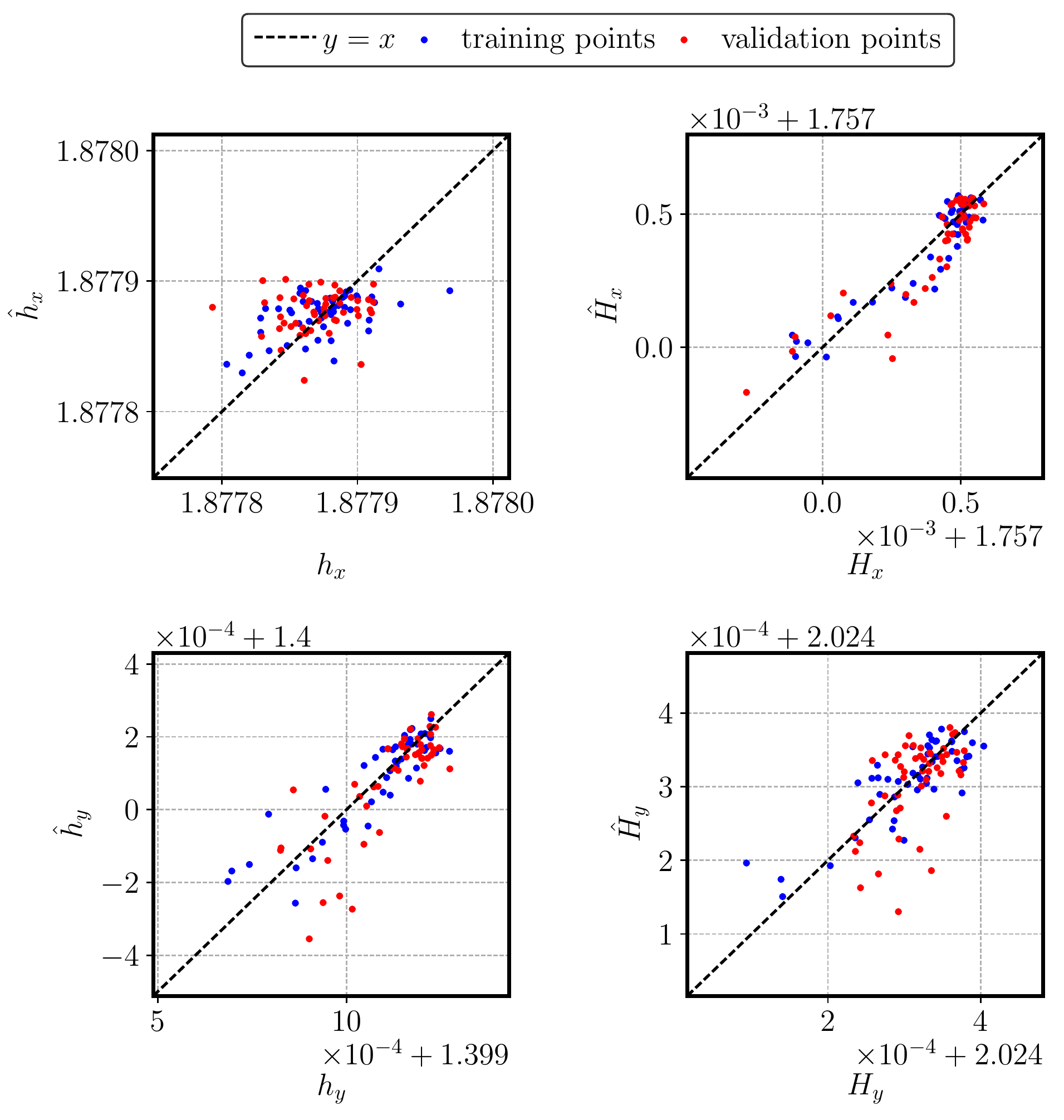}
    \caption{Comparison between the high fidelity ($x$-axis) and PC surrogate model ($y$-axis) at
    \SI[round-mode=places, round-precision=0]{470.79}{\degree}
    of the \ringcyc{} simulation.
    The blue and red dots indicate the training and validation points, respectively. In the best case all points coincide with
    the dashed black line.}
    \label{fig:ring_high_fidelity_vs_surrogate_step_366}
\end{figure}

\begin{figure}[!ht]
    \includegraphics[width=1.0\textwidth]{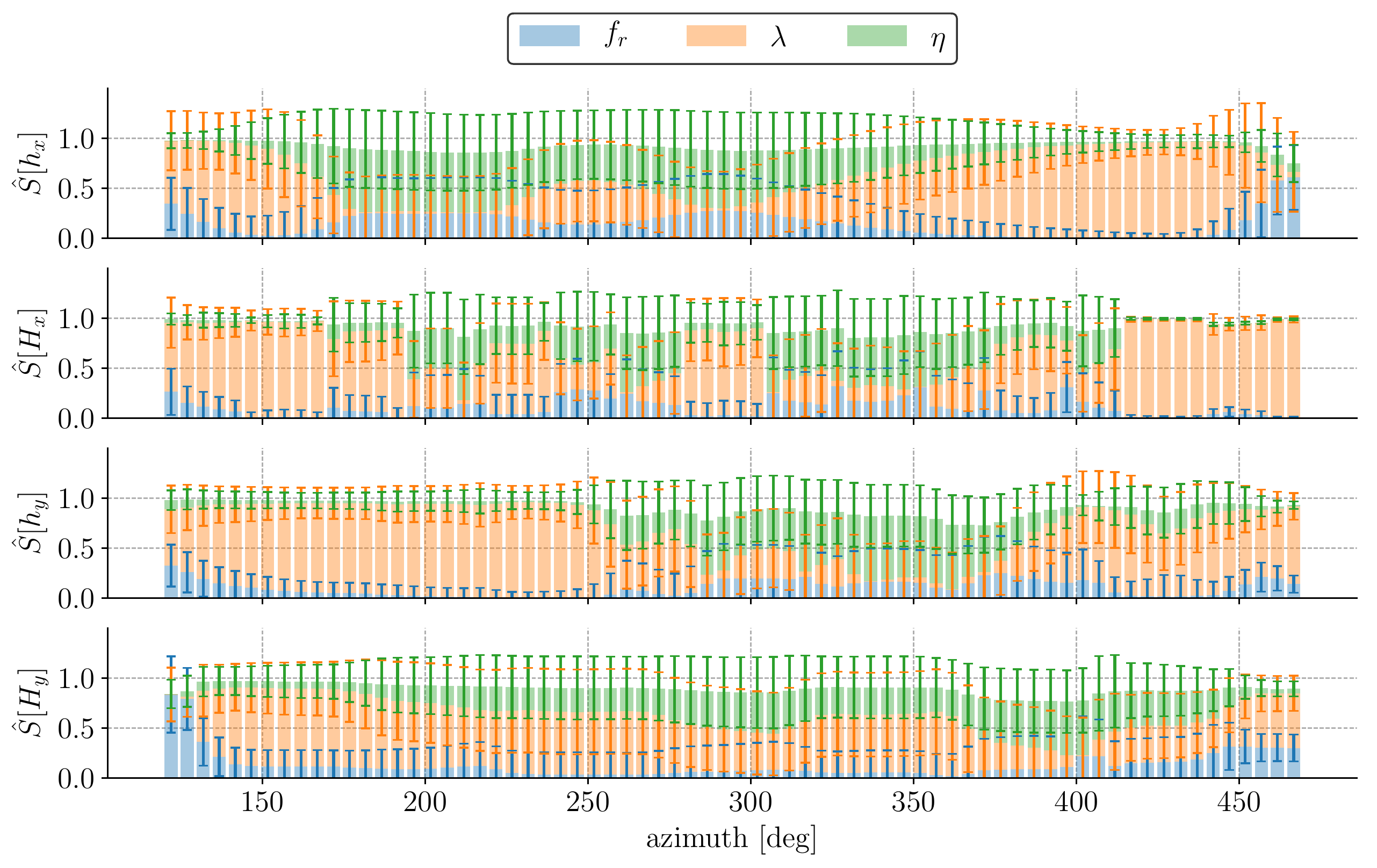}
    \caption{Evolution of the estimated first order Sobol' indices in the \ringcyc.
    The error bars denote the \SI{95}{\percent} bootstrapped $(B=100)$ confidence interval (cf. \Secref{sec:ci_bootstrap}).
    \reviewchange{The main sensitivities are evaluated for the spatial-profile parameters $h_x$, $h_y$ and the phase-space halo parameters $H_x$,
    $H_y$ as defined in \Eqref{eq:profile_parameter} and \Eqref{eq:halo_parameter}, respectively. The bars are coloured with respect
    to the regrid frequency $f_r$, AMR refinement threshold $\lambda$ and energy binning parameter $\eta$. Due to the high uncertainty
    of the sensitivities, no reliable conclusion can be drawn for this machine.}}
    \label{fig:ring_evolution_main_sensitivities}
\end{figure}

In order to study the effect of the input parameters we
evaluate the sensitivities of the halo parameters \Eqref{eq:profile_parameter} and \Eqref{eq:halo_parameter}
with respect to the center bunch of the 5 bunches (cf. \Figref{fig:cyc_sketch}).
The initial kinetic energy of the center bunch in the different models
is approximately \SI{98}{MeV}, \SI{25}{MeV} and \SI{17}{MeV} for the
\ringcyc, \injII{} and DIC, respectively.
As shown in
\Figref{fig:injector2_evolution_training_validation_error}, \Figref{fig:isodar_evolution_training_validation_error} and
\Figref{fig:ring_evolution_training_validation_error}, the relative $l_1$ and $l_2$ errors (cf. \Eqref{eq:rel_l1_error} and
\Eqref{eq:rel_l2_error}) between training and test samples are in good agreement for all cyclotron examples. \reviewchange{A similar
observation is done at a single angle in \Figref{fig:injector2_high_fidelity_vs_surrogate_step_391},
\Figref{fig:isodar_high_fidelity_vs_surrogate_step_107} and \Figref{fig:ring_high_fidelity_vs_surrogate_step_366}}. The average
errors are given in \Tabref{tab:injector2_average_errors_and_average_sensitivities},
\Tabref{tab:isodar_average_errors_and_average_sensitivities} and \Tabref{tab:ring_average_errors_and_average_sensitivities}.
The computation methods OLS, BCS and OMP yield similar results. In case of the \injII, the refinement threshold has more than
\SI{80}{\percent} impact on the halo. The energy binning parameter $\eta$ and regrid frequency $f_r$ play a negligible role.
The increase of the \SI{95}{\percent} bootstrap confidence intervals in \Figref{fig:injector2_evolution_main_sensitivities}
correlates with the decrease of the standard deviation in \Figref{fig:injector2_evolution_mean_std}. It is best observed
for $h_x$ at around \SI{215}{\degree} or $H_x$ between \SI{195}{\degree} and \SI{255}{\degree}. In contrast to the \injII,
the DIC also strongly depends on the regrid frequency. It has an average main sensitivity of approximately \SI{60}{\percent}
for $h_x$. The parameters also exhibit more correlations as observed between the main and total sensitivities (cf.
\Tabref{tab:isodar_average_errors_and_average_sensitivities}). The standard deviations for the DIC are one order of magnitude
smaller than for the \injII, causing the confidence intervals to increase as illustrated in
\Figref{fig:isodar_evolution_main_sensitivities}. This effect is even stronger in the \ringcyc{} where Coulomb's repulsion is
less dominant and the halo parameters are smaller (cf. \Figref{fig:ring_evolution_mean_std}) compared to the \injII. \reviewchange{The standard
deviation is in the order of $\mathcal{O}(10^{-4})$ denoting no significant influence of the input parameters on the model response,
hence, the confidence intervals in \Figref{fig:ring_evolution_main_sensitivities} exhibit large ranges}.
For this reason we can make no reliable statement about the sensitivities for the \ringcyc.
Nevertheless, these findings give rise to computational savings. Due to the small deviations, it is sufficient for the \ringcyc{}
 to select a cheap model. According to \Tabref{tab:cyc_model_timings}, the cheapest model among all $N=100$ samples is
 \num[round-mode=places, round-precision=2]{2.470773495} times faster than the most expensive model.

\begin{table}[!ht]
    \centering
    \begin{tabular}{lS[table-format=3]
                     S[table-format=1.4, round-mode=places, round-precision=4]
                     S[table-format=1.4, round-mode=places, round-precision=4]
                     S[table-format=5, round-mode=places, round-precision=0]}
    \toprule
    model   & \multicolumn{3}{c}{design variables} & {time [\si{s}]} \\ \cmidrule(ll){2-4}
            & {$f_{r}$} & {$\lambda$} & {$\eta$} \\
    \midrule
    \multirow{2}{*}{\ringcyc} & 111 & 0.827205906368997   & 0.0226923203241779   & 7937.66 \\
                              &   3 & 0.502208846849441   & 0.00521209305829928  & 19612.6 \\ \midrule
    \multirow{2}{*}{\injII}   &   4 & 0.645451840951718   & 0.0223818891836287   & 10526.2 \\
                              &   3 & 0.502208846849441   & 0.00446088375246935  & 21422.1 \\ \midrule
    \multirow{2}{*}{DIC}      &  90 & 0.85843651996939729 & 0.015726255173021168 & 9559.72 \\
                              &  74 & 0.59582475626678899 & 0.021640390914407701 & 31709.4 \\ \bottomrule
    \end{tabular}
    \caption{Most expensive and cheapest cyclotron models with respect to runtime among all $N=100$ samples.}
    \label{tab:cyc_model_timings}
\end{table}

%

\begin{table}[!ht]
    \centering
    \sisetup{
             table-format = 1.2,
             table-text-alignment = center,
             round-mode=places,
             round-precision=2,
             scientific-notation=false,
             fixed-exponent=0}
    \begin{tabular}{ccSSSSSSSSSS}
    \toprule
    QoI & method & \multicolumn{2}{c}{$l_1$ error [\si{\percent}]}
                 & \multicolumn{2}{c}{$l_2$ error [\si{\percent}]}
                 & \multicolumn{6}{c}{Sobol' sensitivity indices} \\ \cmidrule(ll){3-4} \cmidrule(ll){5-6} \cmidrule(ll){7-12}
        &        & {train} & {test} & {train} & {test}
                    & {$\hat{S}_{f_{r}}$}
                    & {$\hat{S}_{f_{r}}^T$}
                    & {$\hat{S}_{\lambda}$}
                    & {$\hat{S}_{\lambda}^T$}
                    & {$\hat{S}_{\eta}$}
                    & {$\hat{S}_{\eta}^T$} \\
    \midrule
    \multirow{3}{*}{$h_x$}  & OLS
                            & 0.232821859621 & 0.33031168612
                            & 0.325169056284 & 0.443290907933
                            & 0.0332832164317 & 0.0569199788304
                            & 0.911940255097 & 0.942184764242
                            & 0.0155306434215 & 0.0401440581411 \\
                            & BCS
                            & 0.257338954498 & 0.296255360765
                            & 0.378130065003 & 0.442985128941
                            & 0.0299825140544 & 0.0517329898382
                            & 0.918573246082 & 0.946823461362
                            & 0.015143968892 & 0.037745315176 \\
                            & OMP
                            & 0.243060322065 & 0.317087569515
                            & 0.342042936791 & 0.451375236579
                            & 0.0258241931531 & 0.0449206982465
                            & 0.922471367016 & 0.948979550726
                            & 0.0169018592412 & 0.0409023316175 \\ \midrule
    \multirow{3}{*}{$H_x$}  & OLS
                            & 0.148205498661 & 0.221405898725
                            & 0.205247578735 & 0.291173389144
                            & 0.0789671838295 & 0.161929025011
                            & 0.804487851808 & 0.889096567287
                            & 0.0148920501883 & 0.0506345630965 \\
                            & BCS
                            & 0.16464880834 & 0.196712500816
                            & 0.242120670319 & 0.283876688118
                            & 0.074097134302 & 0.15414123727
                            & 0.814520580464 & 0.896304217515
                            & 0.014347244061 & 0.046597912201 \\
                            & OMP
                            & 0.15440009372 & 0.213103179933
                            & 0.21591081409 & 0.295954037154
                            & 0.0705860621624 & 0.16223979001
                            & 0.803543357493 & 0.898243201997
                            & 0.0147096490569 & 0.0506779392813 \\ \midrule
    \multirow{3}{*}{$h_y$}  & OLS
                            & 0.184056392662 & 0.273807830239
                            & 0.254739486285 & 0.363653831697
                            & 0.0580244501889 & 0.099978128335
                            & 0.87613885097 & 0.91522999667
                            & 0.0117479579936 & 0.0388936771746 \\
                            & BCS
                            & 0.200391486694 & 0.239731290676
                            & 0.299284454823 & 0.353022457333
                            & 0.0508946491652 & 0.094169887657
                            & 0.882519868621 & 0.923567655954
                            & 0.011757881058 & 0.0370985368445 \\
                            & OMP
                            & 0.190515298291 & 0.261395763441
                            & 0.268628573041 & 0.368172349073
                            & 0.04918526033 & 0.0844509816666
                            & 0.888950703942 & 0.923090157184
                            & 0.0135169421099 & 0.0408059547683 \\ \midrule
    \multirow{3}{*}{$H_y$}  & OLS
                            & 0.107423276534 & 0.165587296093
                            & 0.147401293434 & 0.219008577268
                            & 0.0595162996904 & 0.100453925054
                            & 0.878863744688 & 0.918049699489
                            & 0.00870999535568 & 0.0344076692222 \\
                            & BCS
                            & 0.118521563913 & 0.145609358176
                            & 0.174598998574 & 0.212125190429
                            & 0.0576056704506 & 0.0967719840202
                            & 0.883131079495 & 0.921055066205
                            & 0.00830269936831 & 0.0331344434113 \\
                            & OMP
                            & 0.111484167055 & 0.15992809549
                            & 0.155275012987 & 0.222871258764
                            & 0.0503240613482 & 0.0863495234839
                            & 0.889733763289 & 0.927077170279
                            & 0.00880172321226 & 0.0377137583881 \\ \bottomrule
    \end{tabular}
    \caption{Average relative $l_1$ and $l_2$ errors between the high fidelity model and the PC surrogate models
    for the training and validation sets as well as the average
    main and total sensitivities for the \injII. OLS: Ordinary Least Squares;
    BCS: Bayesian Compressive Sensing; OMP: Orthogonal Matching Pursuit.}
    \label{tab:injector2_average_errors_and_average_sensitivities}
\end{table}

\clearpage

\begin{figure}[!ht]
    \includegraphics[width=1.0\textwidth]{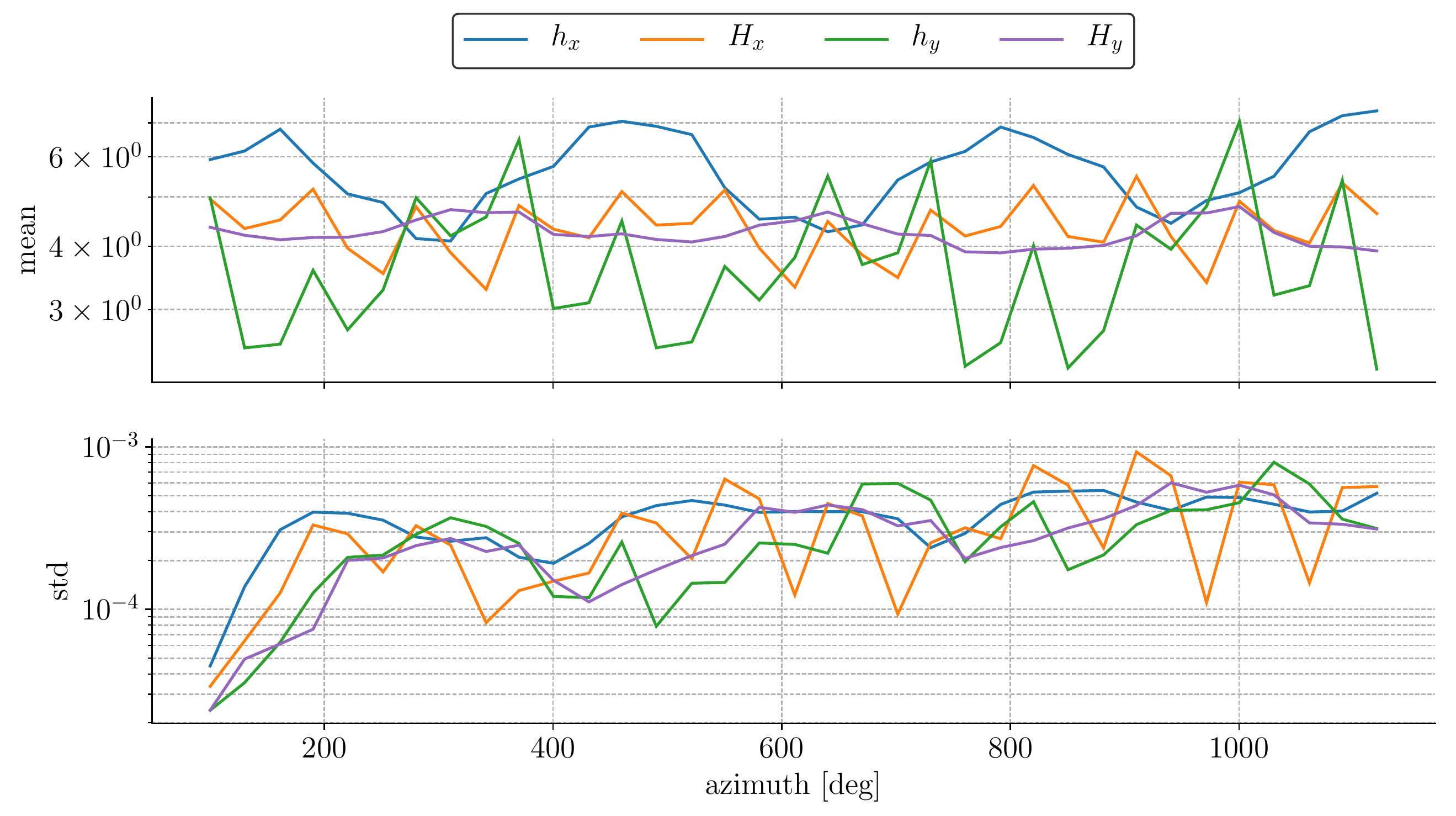}
    \caption{Evolution of the mean and standard deviation (std) \reviewchange{of the spatial-profile parameters $h_x$, $h_y$ and the phase-space
    halo parameters $H_x$, $H_y$ as defined in \Eqref{eq:profile_parameter} and \Eqref{eq:halo_parameter}, respectively. The mean
    of all quantities shows a more or less periodic pattern along the three turns of the \daedalus{} Injector Cyclotron.}}
    \label{fig:isodar_evolution_mean_std}
\end{figure}

%

\begin{table}[!ht]
    \centering
    \sisetup{
             table-format = 1.2,
             table-text-alignment = center,
             round-mode=places,
             round-precision=2,
             scientific-notation=false,
             fixed-exponent=0}
    \begin{tabular}{ccS[fixed-exponent=-2, table-format=1.2,table-omit-exponent]
                      S[fixed-exponent=-2, table-format=1.2,table-omit-exponent]
                      S[fixed-exponent=-2, table-format=1.2,table-omit-exponent]
                      S[fixed-exponent=-2, table-format=1.2,table-omit-exponent]
                    SSSSSS}
    \toprule
    QoI & method & \multicolumn{2}{c}{$l_1$ error [$10^{-2}$\si{\percent}]}
                 & \multicolumn{2}{c}{$l_2$ error [$10^{-2}$\si{\percent}]}
                 & \multicolumn{6}{c}{Sobol' sensitivity indices} \\ \cmidrule(ll){3-4} \cmidrule(ll){5-6} \cmidrule(ll){7-12}
        &        & {train} & {test} & {train} & {test}
                    & {$\hat{S}_{f_{r}}$}
                    & {$\hat{S}_{f_{r}}^T$}
                    & {$\hat{S}_{\lambda}$}
                    & {$\hat{S}_{\lambda}^T$}
                    & {$\hat{S}_{\eta}$}
                    & {$\hat{S}_{\eta}^T$} \\
    \midrule
    \multirow{3}{*}{$h_x$}  & OLS
                            & 0.00294452459308 & 0.00454453446617
                            & 0.00382497656874 & 0.0060531056089
                            & 0.591202348746 & 0.706129435228
                            & 0.265261396713 & 0.334730768575
                            & 0.0164498784036 & 0.08622764077 \\
                            & BCS
                            & 0.00300054398319 & 0.0045809476934
                            & 0.00392282638203 & 0.00608371637621
                            & 0.599446635564 & 0.715895492045
                            & 0.254150529693 & 0.325473002264
                            & 0.0164503306266 & 0.0885854562753 \\
                            & OMP
                            & 0.00299858270809 & 0.00447177298527
                            & 0.00394607740618 & 0.0059976807107
                            & 0.616075462855 & 0.717722883017
                            & 0.264766841987 & 0.319504157065
                            & 0.0115045717905 & 0.0704260832858 \\ \midrule
    \multirow{3}{*}{$H_x$}  & OLS
                            & 0.00214909327872 & 0.00313548392991
                            & 0.00278304203536 & 0.00412199845088
                            & 0.288755932821 & 0.398769861025
                            & 0.565778200161 & 0.669472377271
                            & 0.0211565614829 & 0.056069186005 \\
                            & BCS
                            & 0.00225281704497 & 0.00317312172603
                            & 0.00295908765017 & 0.00415489485727
                            & 0.287237458849 & 0.39687900602
                            & 0.56882368446 & 0.672081853888
                            & 0.0201572959964 & 0.0548225979703 \\
                            & OMP
                            & 0.00234527719379 & 0.00309135619131
                            & 0.00305826106961 & 0.00402446169804
                            & 0.252230120309 & 0.343830126599
                            & 0.644293223368 & 0.725082153242
                            & 0.00964990627773 & 0.0249144702044 \\ \midrule
    \multirow{3}{*}{$h_y$}  & OLS
                            & 0.00340912453798 & 0.00554180951653
                            & 0.00440225059819 & 0.00717790657559
                            & 0.106374659724 & 0.239196354928
                            & 0.67427083939 & 0.764424681492
                            & 0.0695425846108 & 0.14619686837 \\
                            & BCS
                            & 0.00346529287154 & 0.00545991214869
                            & 0.00445686070275 & 0.00705756712603
                            & 0.106200719907 & 0.239379101085
                            & 0.673901118492 & 0.764202010497
                            & 0.0696762110921 & 0.146649757605 \\
                            & OMP
                            & 0.00358543266439 & 0.00537145360217
                            & 0.00457846787561 & 0.00688806865404
                            & 0.101900162904 & 0.221518800453
                            & 0.712196987592 & 0.773835606696
                            & 0.061005594389 & 0.129542847966 \\ \midrule
    \multirow{3}{*}{$H_y$}  & OLS
                            & 0.00242185086071 & 0.00367659521469
                            & 0.0031166010169 & 0.00470898587497
                            & 0.467787967483 & 0.578066105013
                            & 0.388597334787 & 0.477511892662
                            & 0.0188595058026 & 0.0691858132541 \\
                            & BCS
                            & 0.00245061959843 & 0.00380567248849
                            & 0.00317635534799 & 0.00487854031243
                            & 0.467000202266 & 0.577418574766
                            & 0.389491110328 & 0.47840371086
                            & 0.0187748591299 & 0.0689194966902 \\
                            & OMP
                            & 0.00253881501986 & 0.00367825604798
                            & 0.00329921892484 & 0.00466510288548
                            & 0.43663118774 & 0.53451259483
                            & 0.455310318398 & 0.525769056283
                            & 0.00863963331531 & 0.0391372094329 \\ \bottomrule
    \end{tabular}
    \caption{Average relative $l_1$ and $l_2$ errors between the high fidelity model and the PC surrogate models
    for the training and validation sets as well as the average
    main and total sensitivities for the \reviewchange{\daedalus{} Injector Cyclotron}. OLS: Ordinary Least Squares;
    BCS: Bayesian Compressive Sensing; OMP: Orthogonal Matching Pursuit.}
    \label{tab:isodar_average_errors_and_average_sensitivities}
\end{table}

\clearpage

\clearpage

\begin{table}[!ht]
    \centering
    \sisetup{
             table-format = 1.2,
             table-text-alignment = center,
             round-mode=places,
             round-precision=2,
             scientific-notation=false,
             fixed-exponent=0}
    \begin{tabular}{cc
                    S[fixed-exponent=-2, table-format=1.2,table-omit-exponent]
                    S[fixed-exponent=-2, table-format=1.2,table-omit-exponent]
                    S[fixed-exponent=-2, table-format=1.2,table-omit-exponent]
                    S[fixed-exponent=-2, table-format=1.2,table-omit-exponent]
                    SSSSSS}
    \toprule
    QoI & method & \multicolumn{2}{c}{$l_1$ error [$10^{-2}$\si{\percent}]}
                 & \multicolumn{2}{c}{$l_2$ error [$10^{-2}$\si{\percent}]}
                 & \multicolumn{6}{c}{Sobol' sensitivity indices} \\ \cmidrule(ll){3-4} \cmidrule(ll){5-6} \cmidrule(ll){7-12}
        &        & {train} & {test} & {train} & {test}
                    & {$\hat{S}_{f_{r}}$}
                    & {$\hat{S}_{f_{r}}^T$}
                    & {$\hat{S}_{\lambda}$}
                    & {$\hat{S}_{\lambda}^T$}
                    & {$\hat{S}_{\eta}$}
                    & {$\hat{S}_{\eta}^T$} \\
    \midrule
    \multirow{3}{*}{$h_x$}  & OLS
                            & 0.00846216916924 & 0.00780691832899
                            & 0.0112718792964 & 0.0101213256643
                            & 0.149564698832 & 0.193514991919
                            & 0.502307714343 & 0.544650075988
                            & 0.272360345065 & 0.337610396784  \\
                            & BCS
                            & 0.00919078595462 & 0.00801577367849
                            & 0.0124144675534 & 0.0105525570512
                            & 0.150116844068 & 0.194318598477
                            & 0.500931023206 & 0.542756695765
                            & 0.273150475296 & 0.338735998094 \\
                            & OMP
                            & 0.00864989463822 & 0.00783857051338
                            & 0.0114268331227 & 0.0100799510237
                            & 0.136369694631 & 0.161875388105
                            & 0.50041232181 & 0.508579420757
                            & 0.336377012981 & 0.356386161717 \\ \midrule
    \multirow{3}{*}{$H_x$}  & OLS
                            & 0.00472163071874 & 0.00446317588695
                            & 0.00623452487337 & 0.0057898182172
                            & 0.10665644507 & 0.16291857749
                            & 0.592377759679 & 0.620656635265
                            & 0.223437409262 & 0.293955700649 \\
                            & BCS
                            & 0.00598221766646 & 0.0050136739088
                            & 0.00785911016308 & 0.00645295422226
                            & 0.105142411058 & 0.16207443971
                            & 0.593584665138 & 0.62188990822
                            & 0.222980027494 & 0.294331253212 \\
                            & OMP
                            & 0.00484163772208 & 0.00448570138594
                            & 0.00635977690094 & 0.00569262785951
                            & 0.0665966578481 & 0.127171361122
                            & 0.634645132509 & 0.636749596708
                            & 0.238183506369 & 0.296653745444 \\ \midrule
    \multirow{3}{*}{$h_y$}  & OLS
                            & 0.00800924092879 & 0.00756816512383
                            & 0.0105079531783 & 0.00980142994402
                            & 0.102610063899 & 0.185343809241
                            & 0.59730014385 & 0.631890288136
                            & 0.204240535349 & 0.278616854186 \\
                            & BCS
                            & 0.00956054898038 & 0.00804844925812
                            & 0.0127924483619 & 0.0106396836777
                            & 0.102418438962 & 0.185249557747
                            & 0.597505793032 & 0.632039321571
                            & 0.204164696684 & 0.278623934309 \\
                            & OMP
                            & 0.0081405734972 & 0.00768991430657
                            & 0.0106663496925 & 0.00982044716154
                            & 0.0884869581729 & 0.165845172023
                            & 0.588217322146 & 0.606066207961
                            & 0.236689129702 & 0.314695209995 \\ \midrule
    \multirow{3}{*}{$H_y$}  & OLS
                            & 0.00417376930154 & 0.00385312494114
                            & 0.00551769591752 & 0.0049722686239
                            & 0.129490345249 & 0.219391367739
                            & 0.510831201465 & 0.554174815745
                            & 0.250473440194 & 0.335749781728 \\
                            & BCS
                            & 0.00474133572541 & 0.00387923485479
                            & 0.00623022834555 & 0.00493995334084
                            & 0.134308726874 & 0.226938365147
                            & 0.501128720261 & 0.546888520138
                            & 0.250512277877 & 0.340311450774 \\
                            & OMP
                            & 0.00428088003666 & 0.00404044740393
                            & 0.00563724126405 & 0.0049916410625
                            & 0.0976651610215 & 0.226235305757
                            & 0.52368363599 & 0.540477683772
                            & 0.244576318139 & 0.367361895321 \\ \bottomrule
    \end{tabular}
    \caption{Average relative $l_1$ and $l_2$ errors between the high fidelity model and the PC surrogate models
    for the training and validation sets as well as the average
    main and total sensitivities for the \ringcyc. OLS: Ordinary Least Squares;
    BCS: Bayesian Compressive Sensing; OMP: Orthogonal Matching Pursuit.}
    \label{tab:ring_average_errors_and_average_sensitivities}
\end{table}

\clearpage

\subsection{RF Electron Gun Model}
\begin{figure}[!ht]
    \includegraphics[width=1.0\textwidth]{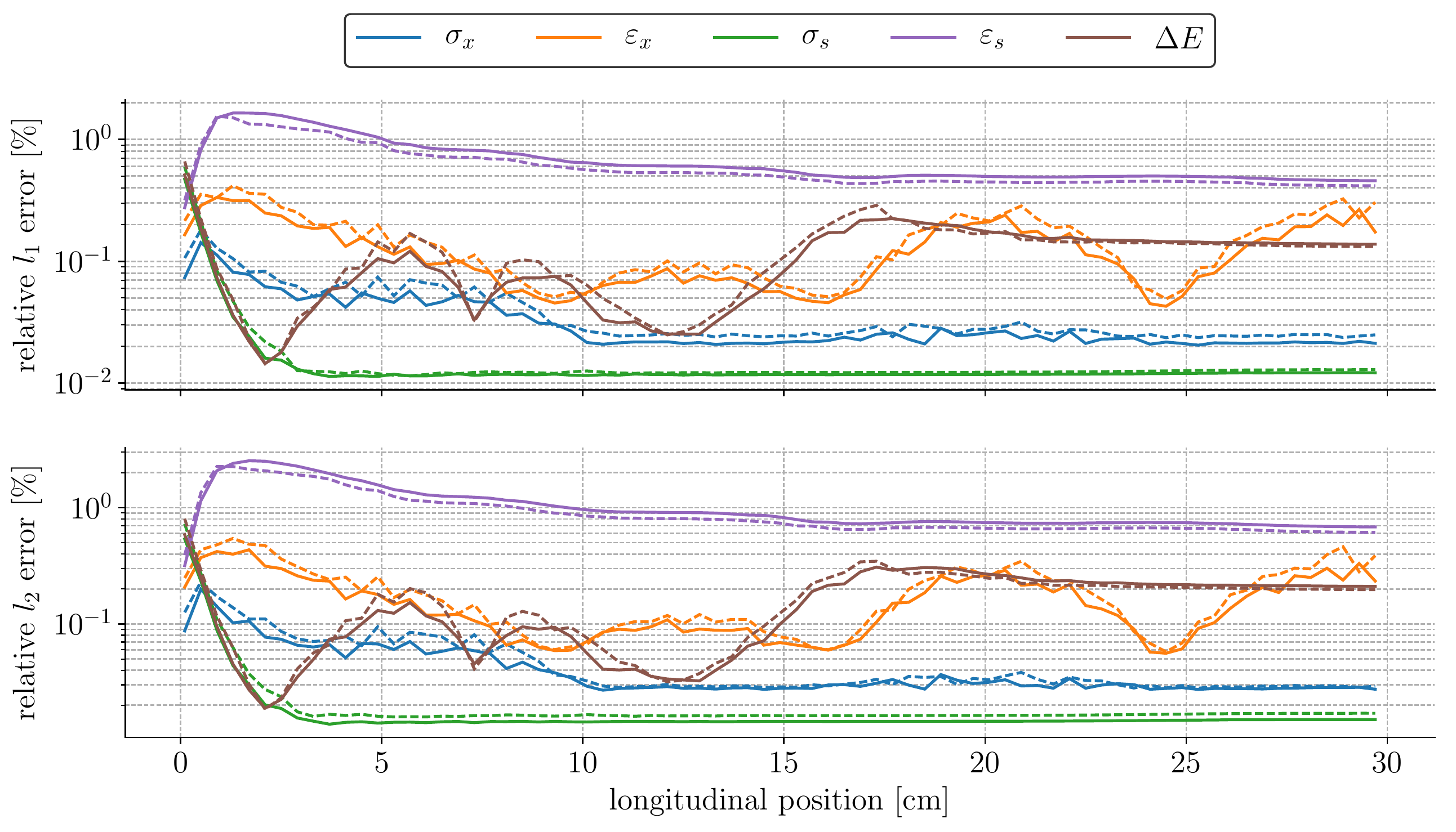}
    \caption{Evolution of the relative $l_2$ and $l_1$ error \reviewchange{between the surrogate and the true model
    of the AWA rf electron gun. The full lines are the errors to the
    surrogate model obtained with the training set and the dashed lines are the errors to the surrogate model
    computed with the validation set. For each quantity, the dashed and full lines are close to each other, indicating
    no overfitting of the surrogate model.}}
    \label{fig:awa_evolution_training_validation_error}
\end{figure}

\begin{figure}[!ht]
    \includegraphics[width=1.0\textwidth]{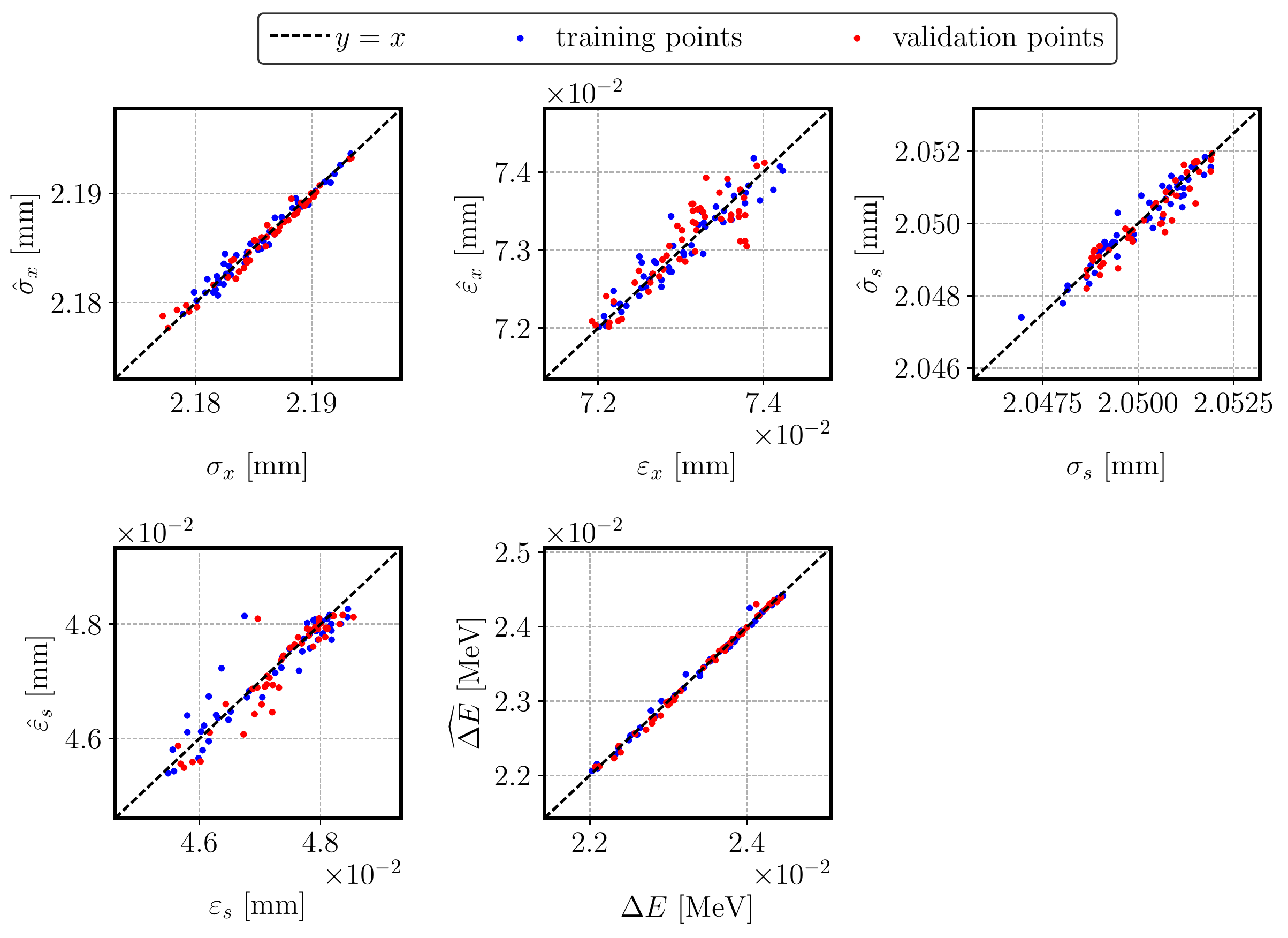}
    \caption{Comparison between the high fidelity ($x$-axis) and PC surrogate model ($y$-axis) at the exit of
    the rf electron gun model of the AWA, i.e. $s \approx \SI{30}{cm}$.
    The blue and red dots indicate the training and validation points, respectively. In the best case all points coincide with
    the dashed black line.}
    \label{fig:high_fidelity_vs_surrogate_step_299}
\end{figure}

In order to approximate the high fidelity model we use PC surrogate models of second order where the BCS method uses a tolerance of
$\varepsilon = 10^{-9}$ and the OMP method is stopped once 7 non-zero coefficients are found.
In \Figref{fig:awa_evolution_training_validation_error} are the relative $l_2$ and $l_1$ errors evaluated along the rf electron gun model.
The mean errors are summarized in \Tabref{tab:awa_average_errors_and_average_sensitivities}. It shows that
the $l_1$ and $l_2$ errors on the test and training points match with an absolute difference of $\mathcal{O}(10^{-2})$ and
$\mathcal{O}(10^{-1})$, respectively. An example of a comparison between the PC surrogate and high fidelity model
is illustrated in \Figref{fig:high_fidelity_vs_surrogate_step_299}.

\begin{figure}[!ht]
    \includegraphics[width=1.0\textwidth]{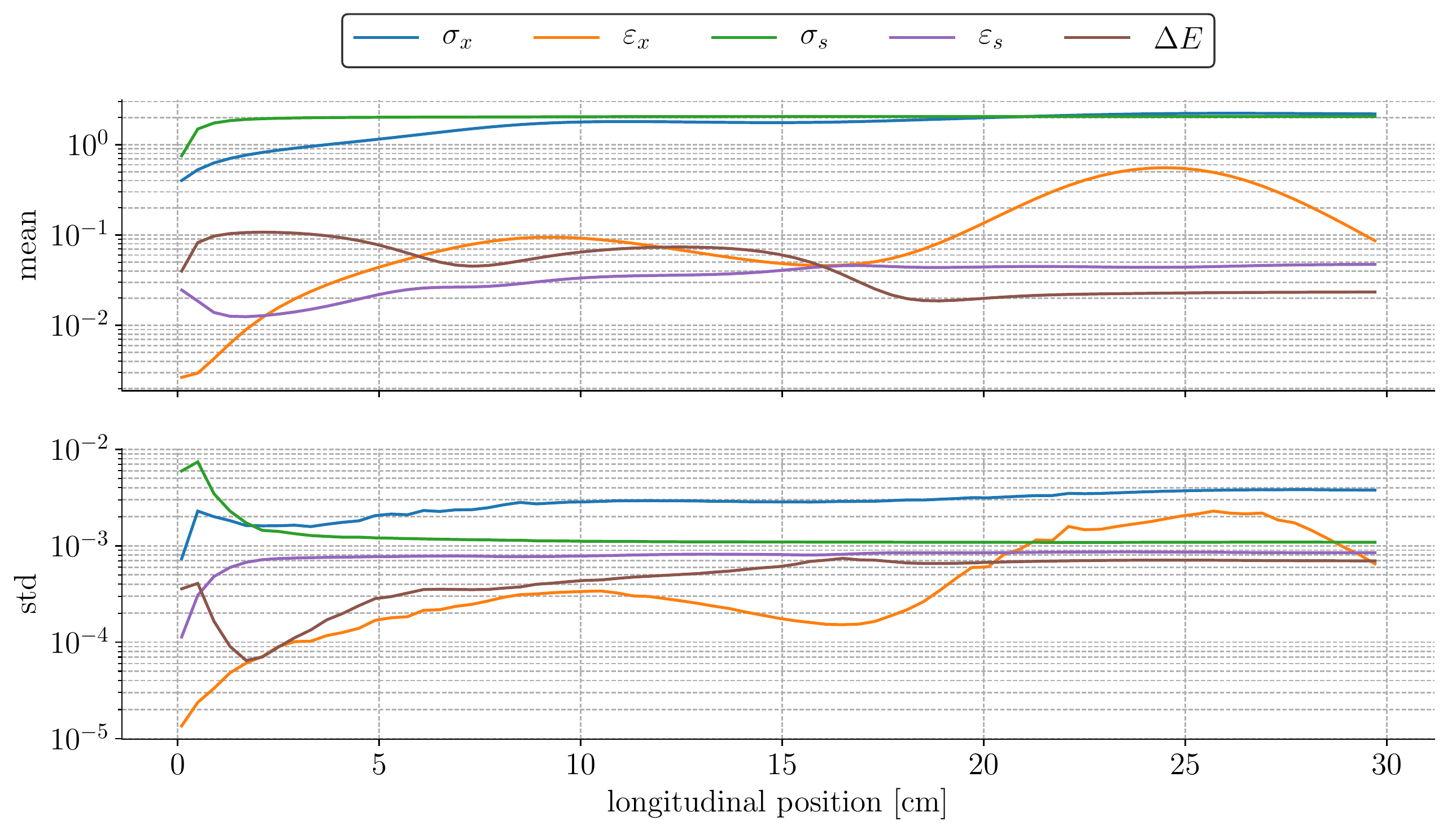}
    \caption{Evolution of the mean and standard deviation (std) \reviewchange{of the energy spread $\Delta E$, the projected
    emittances $\varepsilon_x$, $\varepsilon_s$ and the rms beam sizes $\sigma_x$, $\sigma_s$ for rf electron gun model of the AWA.}}
    \label{fig:awa_evolution_mean_std}
\end{figure}

\begin{figure}[!ht]
    \includegraphics[width=1.0\textwidth]{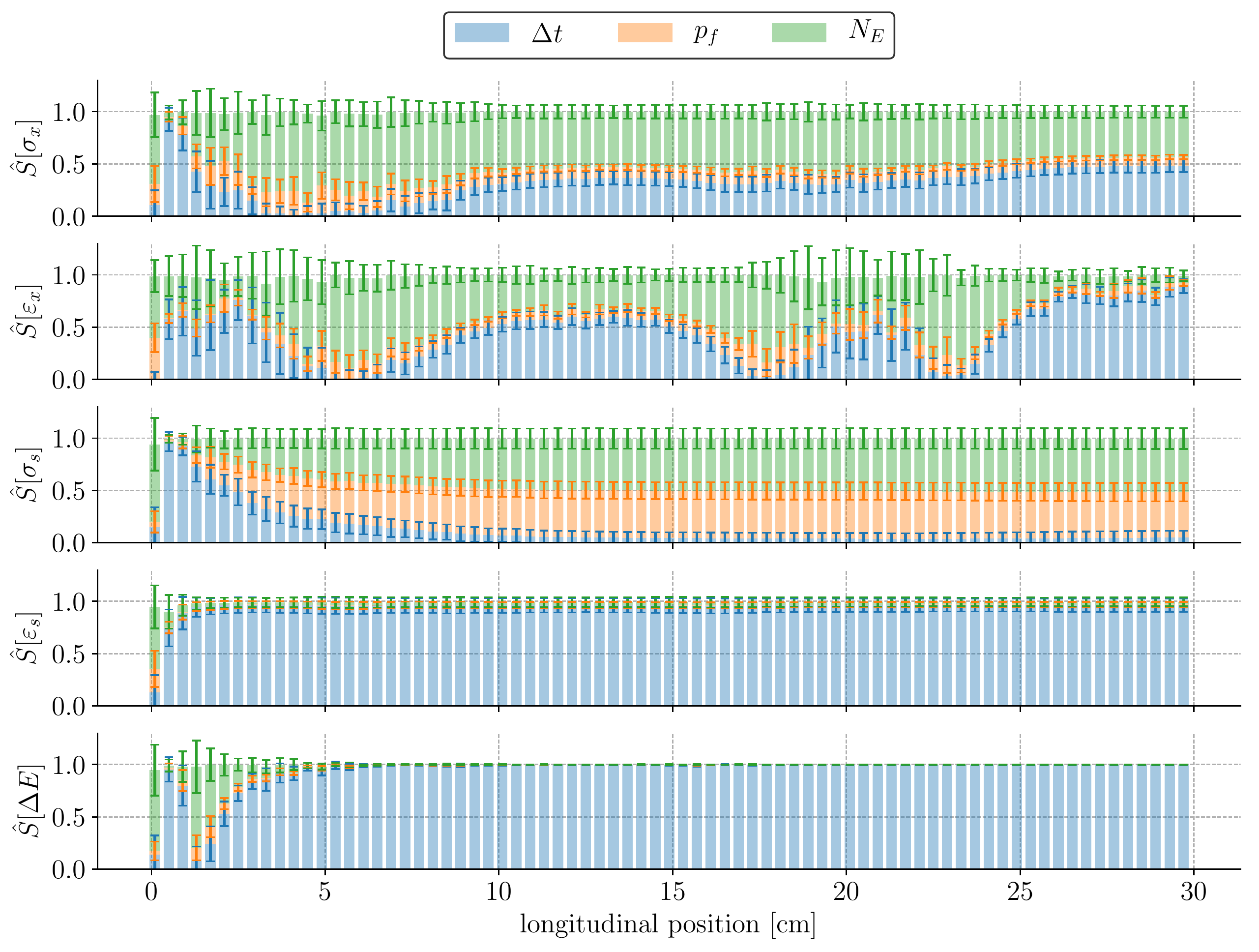}
    \caption{Evolution of the estimated first order Sobol' indices in the rf electron gun model of the AWA.
    The error bars denote the \SI{95}{\percent} bootstrapped $(B=100)$ confidence interval (cf. \Secref{sec:ci_bootstrap}).
    \reviewchange{The main sensitivities are evaluated for the energy spread $\Delta E$, the projected emittances $\varepsilon_x$, $\varepsilon_s$
    and the rms beam sizes $\sigma_x$, $\sigma_s$. The bars are coloured with respect to the time step $\Delta t$, particle
    multiplication factor $p_f$ and the number of energy bins $N_E$. The impact of the parameters on the quantities in longitudinal
    direction converges.}}
    \label{fig:awa_evolution_main_sensitivities}
\end{figure}

The first order Sobol' indices and their \SI{95}{\percent} bootstrapped confidence intervals are illustrated in
\Figref{fig:awa_evolution_main_sensitivities}. Except to the sensitivities of the horizontal projected emittance $\varepsilon_x$,
we observe a convergence of the model parameter influences. The energy spread $\Delta E$ and the projected emittance $\varepsilon_s$
strongly depend on the time step ($\hat{S}[\varepsilon_s], \hat{S}[\Delta E] > 0.90$). This high influence is due
to the momentum component in their definitions (cf. \Eqref{eq:energy_spread} and \Eqref{eq:proj_emit}) and the fact that
\reviewchange{the smaller the time step, the better the process of acceleration (i.e. the evolution of the momentum) is resolved}.
The rms beam
size in longitudinal direction is dominated by the energy binning ($\hat{S}[N_{E}]\approx 0.45$)
and the particle multiplication factor ($\hat{S}[p_{f}]\approx 0.41$). While a higher $p_{f}$ value improves the statistics of the
beam size and reduces the numerical noise of PIC, the energy binning is coupled with Coulomb's repulsion that affects the beam size.
In transverse direction, $N_E$ and $\Delta t$ are important instead.
The convergence of the relative errors is correlated with the convergence of the variances of the quantities of interest as
observed in \Figref{fig:awa_evolution_mean_std}. The model might therefore be improved with an adaptive time stepping scheme that
addresses this effect. The cheapest AWA rf electron gun model, i.e. $\Delta t=\SI{1}{ps}$,  $N_E =2$ and $p_{f} =1$,  is
\num[round-mode=places, round-precision=0]{15.237717908} times faster than the most expensive model which has
$\Delta t=\SI{0.1}{ps}$, $N_{E} =10$ and $p_{f} = 5$.

\begin{table}[!ht]
    \centering
    \sisetup{
             table-format = 1.2,
             table-text-alignment = center,
             round-mode=places,
             round-precision=2,
             scientific-notation=false,
             fixed-exponent=0}
    \begin{tabular}{ccSSSSSSSSSS}
    \toprule
    QoI & method & \multicolumn{2}{c}{$l_1$ error [\si{\percent}]}
                 & \multicolumn{2}{c}{$l_2$ error [\si{\percent}]}
                 & \multicolumn{6}{c}{Sobol' sensitivity indices} \\ \cmidrule(ll){3-4} \cmidrule(ll){5-6} \cmidrule(ll){7-12}
        &        & {train} & {test} & {train} & {test}
                    & {$\hat{S}_{\Delta t}$}
                    & {$\hat{S}_{\Delta t}^T$}
                    & {$\hat{S}_{p_{f}}$}
                    & {$\hat{S}_{p_{f}}^T$}
                    & {$\hat{S}_{N_{E}}$}
                    & {$\hat{S}_{N_{E}}^T$} \\
    \midrule
    \multirow{3}{*}{$\sigma_x$}      & OLS
                                     & 0.0337561152513 & 0.0401158483048
                                     & 0.0432846442772 & 0.0495257904284
                                     & 0.317231191336 & 0.321293769766
                                     & 0.118960256283 & 0.121559097556
                                     & 0.558751198852 & 0.562204486889 \\
                                     & BCS
                                     & 0.0375100314081 & 0.0424948254229
                                     & 0.0476949355223 & 0.0526338986125
                                     & 0.317231245341 & 0.321293723198
                                     & 0.118960205139 & 0.12155896272
                                     & 0.558751322113 & 0.562204542181 \\
                                     & OMP
                                     & 0.0342345868394 & 0.0399533763696
                                     & 0.0439747003974 & 0.0489441009452
                                     & 0.320301605606 & 0.323015745883
                                     & 0.121658066876 & 0.122919744397
                                     & 0.554655554331 & 0.557449282905 \\ \midrule
    \multirow{3}{*}{$\varepsilon_x$} & OLS
                                     & 0.128644137131 & 0.156742084063
                                     & 0.163378794155 & 0.198031850182
                                     & 0.433188470507 & 0.443277695714
                                     & 0.0926983184835 & 0.0994173378344
                                     & 0.461647098119 & 0.469771079479 \\
                                     & BCS
                                     & 0.136254491327 & 0.154709687587
                                     & 0.173074898196 & 0.194918196446
                                     & 0.433223011013 & 0.443176165331
                                     & 0.0927336347894 & 0.0993416471318
                                     & 0.461720910044 & 0.469804631831 \\
                                     & OMP
                                     & 0.131395584226 & 0.156920045831
                                     & 0.165841989519 & 0.197615124755
                                     & 0.439830037115 & 0.44817551492
                                     & 0.095892210423 & 0.101166452233
                                     & 0.45384567293  & 0.461090112379\\ \midrule
    \multirow{3}{*}{$\sigma_s$}      & OLS
                                     & 0.0214522532739 & 0.0245907442353
                                     & 0.0262425576332 & 0.031754088835
                                     & 0.131912434386 & 0.137392208661
                                     & 0.408145899229 & 0.410322235549
                                     & 0.454068394908 & 0.45815883262 \\
                                     & BCS
                                     & 0.0219714105743 & 0.026853871754
                                     & 0.0282785352843 & 0.0342361889483
                                     & 0.131912613948 & 0.1373923261
                                     & 0.408145520001 & 0.410321790665
                                     & 0.454068679868 & 0.458159075069 \\
                                     & OMP
                                     & 0.0217712866895 & 0.0249719236287
                                     & 0.0267877666371 & 0.0317081468075
                                     & 0.13262226646 & 0.134354362331
                                     & 0.387640274979 & 0.388378729587
                                     & 0.477944909877 & 0.479059456768 \\ \midrule
    \multirow{3}{*}{$\varepsilon_s$} & OLS
                                     & 0.694447125841 & 0.616410825971
                                     & 1.04129198063 &  0.929607880117
                                     & 0.9463566143 & 0.956278013311
                                     & 0.00775800999687 & 0.00851918175843
                                     & 0.0358731907096 & 0.0452149899282 \\
                                     & BCS
                                     & 0.925594370168 & 0.860688730597
                                     & 1.21918810486 & 1.12372570859
                                     & 0.946581221681 & 0.956227460319
                                     & 0.00785308764483 & 0.00863777105423
                                     & 0.0358285545313 & 0.0448719047731 \\
                                     & OMP
                                     & 0.694242824635 & 0.625872178206
                                     & 1.0446660571 & 0.94751591923
                                     & 0.946914401994 & 0.957263379485
                                     & 0.00761692592026 & 0.00809084361639
                                     & 0.035119694595 & 0.04499475439 \\ \midrule
    \multirow{3}{*}{$\Delta E$}      & OLS
                                     & 0.114272864138 & 0.123998435711
                                     & 0.159888329298 & 0.167836845715
                                     & 0.942337717858 & 0.944768577303
                                     & 0.00871512090948 & 0.0103529514648
                                     & 0.0461879005668 & 0.0476377321484 \\
                                     & BCS
                                     & 0.143833321344 & 0.154745425134
                                     & 0.184751110548 & 0.194996806352
                                     & 0.942386571447 & 0.944799182955
                                     & 0.00868920045955 & 0.0103310916576
                                     & 0.0461836431045 & 0.0476103106015 \\
                                     & OMP
                                     & 0.114617904809 & 0.12285284897
                                     & 0.16150366619 & 0.165561315705
                                     & 0.942284146628 & 0.94427259405
                                     & 0.0084462538913 & 0.00956720479345
                                     & 0.0470915441137 &  0.048338256524 \\ \bottomrule
    \end{tabular}
    \caption{Average relative $l_1$ and $l_2$ errors between the high fidelity model and the PC surrogate models
    for the training and validation sets as well as the average
    main and total sensitivities for the rf electron gun model of the AWA. OLS: Ordinary Least Squares;
    BCS: Bayesian Compressive Sensing; OMP: Orthogonal Matching Pursuit.}
    \label{tab:awa_average_errors_and_average_sensitivities}
\end{table}

\clearpage

\section{Conclusions}
\label{sec:conclusions}
In this paper we discussed uncertainty quantification based on polynomial chaos expansion and  \reviewchange{gave a brief introduction}{} to four
numerical methods to compute the polynomial coefficients. \reviewchange{The choice of the method depends on the problem and its dimension.
While the projection method is the most accurate, the number of high-fidelity evaluations grows exponentially with the dimension which
is not the case for the other presented methods. Bayesian compressive sensing and matching orthogonal pursuit favour sparse solutions
by the selection of the most important contributions. The least squares method solves a linear system which may fail in case the
matrix is ill-conditioned which happens when input dimensions can only take a few discrete values. This can be the case with
integer input, for example, when specifying the number of AMR levels or the grid sizes for the Poisson solver.

Beside a cheap surrogate model that mimics the high fidelity model, polynomial chaos based uncertainty quantification has the
additional benefit to
easily evaluate the Sobol' sensitivity indices}. As demonstrated in this paper, this technique is not only suitable to gain knowledge
about the sensitivity of physical parameters (e.g. initial beam properties) on the quantities of \reviewchange{interest}{}
but also numerical parameters of computer codes. Since some tested numerical parameters might be limited to
integers, the projection method to obtain the polynomial coefficients is, however, not applicable. Instead, regression-based
methods, Bayesian Compressive Sensing or Orthogonal Matching Pursuit and others have to be applied. A further difficulty with
numerical parameters is a fair random sampling. In some cases (cf. \Figref{fig:binning}) a
straightforward, uniform sampling of the parameter yields biased input data and, hence, may induce wrong conclusions. To
circumvent, we perform a stratified sampling that guarantees a well-balanced distribution.

The sensitivity studies of the three high intensity cyclotrons show that the sensitivity results are different among accelerators
of the same type. While the AMR threshold is the most important parameter in the \injII{}
with a sensitivity of about \SI{90}{\percent}, the regrid frequency is relevant in the \daedalus{} Injectory Cyclotron (DIC), too.
\reviewchange{Large bootstrap confidence intervals for the Sobol' indices indicate a failure of the analysis since the contributed variation of the
model response is rather due to noise than the tested input parameters}.
In such a case no reliable statements based on the
sensitivity estimates can be done.
In contrast to our intuition the standard deviation of the halo parameters remain pretty constant throughout one turn in
the \ringcyc{} and \injII{} and the considered three turns in the DIC.
Nevertheless, these findings give rise to computational savings. Without losing significantly on accuracy
(cf. \Figref{fig:injector2_evolution_mean_std}, \Figref{fig:isodar_evolution_mean_std}
and \Figref{fig:ring_evolution_mean_std}), energy binning can be totally switched off for these
cyclotrons. In addition, this reduces the amount of AMR hierarchy updates which reduces the time-to-solution even further since
the operators of the adaptive multigrid solver to solve Poisson's equation do not need to be set up in every time step. To illustrate
this, we take the benchmark example in \cite{FREY2019106912} that solves Poisson's equation
100 times using a three level AMR hierarchy with a base level of $576^3$ grid points. The benchmark running on \num{14400} CPU (Central Processing Unit) cores shows that the matrix setup
due to AMR regriding takes up \SI{42.15}{\percent} computing time. A reduction of the regrid frequency by
a factor 10 yields a speedup of \num[round-mode=places, round-precision=2]{7.097821254} in the matrix setup timing. In our UQ
samples, the speedup between the cheapest and most expensive model is at least \num{2.0} and
at most \num{3.3}.

Another interesting case we have studied is the rf electron gun model of the AWA. Relevant parameters for this model are the energy binning
$N_{E}$ and the time step $\Delta t$. The particle multiplication factor $p_{f}$, that basically controls the number of particles per
grid cell, is only important for the longitudinal beam size. Although $N_{E}$ and $p_{f}$ have together an average main sensitivity of
\SI[round-mode=places, round-precision=0]{86.22} {\percent}, $\Delta t$ is the dominating parameter close to the cathode. An
adaptive energy binning and time step scheme that is based on Sobol' sensitivity indices is therefore a possible future enhancement
to reduce the time-to-solution for a target accuracy.

\section{Acknowledgments}
The authors thank the developers of \uqtk, especially B.~Debusschere and
K.~Sargsyan. The computer resources were provided by the Swiss National Supercomputing Centre (CSCS),
Paul Scherrer Institut (PSI) and the Laboratory Computing Resource Center at the Argonne National Laboratory (ANL).
This project is funded by the Swiss National Science Foundation (SNSF) under contract number 200021\_159936.

\bibliographystyle{phcpc}

\begin{thebibliography}{10}

\bibitem{SOBOL2001271}
Sobol', I.~M.,
\newblock Mathematics and Computers in Simulation {\bf 55} (2001) 271,
\newblock The Second IMACS Seminar on Monte Carlo Methods.

\bibitem{doi:10.1146/annurev.fluid.010908.165248}
Najm, H.~N.,
\newblock Annual Review of Fluid Mechanics {\bf 41} (2009) 35.

\bibitem{doi:10.1088/1364-7830/8/3/010}
Reagan, M.~T. et~al.,
\newblock Combustion Theory and Modelling {\bf 8} (2004) 607.

\bibitem{doi:10.1002/cnm.2755}
Eck, V.~G. et~al.,
\newblock International Journal for Numerical Methods in Biomedical Engineering
  {\bf 32} (2016) e02755.

\bibitem{doi:10.1137/16M1061928}
Adelmann, A.,
\newblock SIAM/ASA Journal on Uncertainty Quantification {\bf 7} (2019) 383.

\bibitem{doi:10.1061/9780784413609.257}
Marelli, S. and Sudret, B.,
\newblock {UQL}ab: {A} {F}ramework for {U}ncertainty {Q}uantification in
  {M}atlab,
\newblock in {\em Vulnerability, Uncertainty, and Risk}, Second International
  Conference on Vulnerability and Risk Analysis and Management (ICVRAM) and the
  Sixth International Symposium on Uncertainty, Modeling, and Analysis (ISUMA),
  pages 2554--2563, 2014.

\bibitem{FEINBERG201546}
Feinberg, J. and Langtangen, H.~P.,
\newblock Journal of Computational Science {\bf 11} (2015) 46 .

\bibitem{dakota}
Adams, B.~M. et~al.,
\newblock Dakota, a multilevel parallel object-oriented framework for design
  optimization, parameter estimation, uncertainty quantification, and
  sensitivity analysis: Version 6.0 user’s manual, 2014,
\newblock Sandia Technical Report SAND2014-4633, Updated November 2015 (Version
  6.3).

\bibitem{SUDRET2008964}
Sudret, B.,
\newblock Reliability Engineering \& System Safety {\bf 93} (2008) 964 ,
\newblock Bayesian Networks in Dependability.

\bibitem{PhysRevSTAB.9.064402}
Fubiani, G., Qiang, J., Esarey, E., Leemans, W.~P., and Dugan, G.,
\newblock Phys. Rev. ST Accel. Beams {\bf 9} (2006) 064402.

\bibitem{PhysRevSTAB.13.064201}
Yang, J.~J., Adelmann, A., Humbel, M., Seidel, M., and Zhang, T.~J.,
\newblock Phys. Rev. ST Accel. Beams {\bf 13} (2010) 064201.

\bibitem{2019arXiv190506654A}
{Adelmann}, A. et~al.,
\newblock arXiv e-prints  (2019) arXiv:1905.06654.

\bibitem{FREY2019106912}
Frey, M., Adelmann, A., and Locans, U.,
\newblock Computer Physics Communications  (2019) 106912.

\bibitem{BCS_4524050}
{Ji}, S., {Xue}, Y., and {Carin}, L.,
\newblock IEEE Transactions on Signal Processing {\bf 56} (2008) 2346.

\bibitem{BCS_5256324}
{Babacan}, S.~D., {Molina}, R., and {Katsaggelos}, A.~K.,
\newblock IEEE Transactions on Image Processing {\bf 19} (2010) 53.

\bibitem{doi:10.1137/S1064827503427741}
Debusschere, B. et~al.,
\newblock SIAM Journal on Scientific Computing {\bf 26} (2004) 698.

\bibitem{Debusschere2017}
Debusschere, B., Sargsyan, K., Safta, C., and Chowdhary, K.,
\newblock {\em Uncertainty Quantification Toolkit (UQTk)}, pages 1807--1827,
\newblock Springer International Publishing, Cham, 2017.

\bibitem{258082}
{Mallat}, S.~G. and {Zhifeng Zhang},
\newblock IEEE Transactions on Signal Processing {\bf 41} (1993) 3397.

\bibitem{342465}
{Pati}, Y.~C., {Rezaiifar}, R., and {Krishnaprasad}, P.~S.,
\newblock Orthogonal matching pursuit: recursive function approximation with
  applications to wavelet decomposition,
\newblock in {\em Proceedings of 27th Asilomar Conference on Signals, Systems
  and Computers}, pages 40--44 vol.1, 1993.

\bibitem{scikit-learn}
Pedregosa, F. et~al.,
\newblock Journal of Machine Learning Research {\bf 12} (2011) 2825.

\bibitem{sklearn_api}
Buitinck, L. et~al.,
\newblock {API} design for machine learning software: experiences from the
  scikit-learn project,
\newblock in {\em ECML PKDD Workshop: Languages for Data Mining and Machine
  Learning}, pages 108--122, 2013.

\bibitem{MORRISSEY200390}
Morrissey, D.~J., Sherrill, B.~M., Steiner, M., Stolz, A., and Wiedenhoever,
  I.,
\newblock Nuclear Instruments and Methods in Physics Research Section B: Beam
  Interactions with Materials and Atoms {\bf 204} (2003) 90 ,
\newblock 14th International Conference on Electromagnetic Isotope Separators
  and Techniques Related to their Applications.

\bibitem{Alonso2019}
Alonso, J.~R., Barlow, R., Conrad, J.~M., and Waites, L.~H.,
\newblock Nature Reviews Physics {\bf 1} (2019) 533.

\bibitem{FISCHER19971202}
Fischer, W.~E.,
\newblock Physica B: Condensed Matter {\bf 234-236} (1997) 1202 ,
\newblock Proceedings of the First European Conference on Neutron Scattering.

\bibitem{MCCARTHY199735}
McCarthy, D.~W. et~al.,
\newblock Nuclear Medicine and Biology {\bf 24} (1997) 35 .

\bibitem{WEBER2005401}
Weber, D.~C. et~al.,
\newblock International Journal of Radiation Oncology Biology Physics {\bf 63}
  (2005) 401 .

\bibitem{doi:10.1063/1.4802375}
Alonso, J.~R. and {DAE$\delta$ALUS Collaboration},
\newblock AIP Conference Proceedings {\bf 1525} (2013) 480.

\bibitem{doi:10.1063/1.4826879}
Adelmann, A. and {DAE$\delta$ALUS Collaboration},
\newblock AIP Conference Proceedings {\bf 1560} (2013) 709.

\bibitem{gai_power_jing_2012}
Gai, W., Power, J.~G., and Jing, C.,
\newblock Journal of Plasma Physics {\bf 78} (2012) 339–345.

\bibitem{JING201872}
Jing, C. et~al.,
\newblock Nuclear Instruments and Methods in Physics Research Section A:
  Accelerators, Spectrometers, Detectors and Associated Equipment {\bf 898}
  (2018) 72 .

\bibitem{Conde:IPAC2017-WEPAB132}
\reviewchange{M. E. Conde}{} et~al.,
\newblock {R}esearch {P}rogram and {R}ecent {R}esults at the {A}rgonne
  {W}akefield {A}ccelerator {F}acility ({AWA}),
\newblock in {\em Proc. of International Particle Accelerator Conference
  (IPAC'17), Copenhagen, Denmark, 14-19 May, 2017}, number~8 in International
  Particle Accelerator Conference, pages 2885--2887, Geneva, Switzerland, 2017,
  JACoW.

\bibitem{Wangler}
Wangler, T.~P. and Crandall, K.~R.,
\newblock Beam halo in proton linac beams,
\newblock Number~20 in International Linac Conference, 2000.

\bibitem{PhysRevSTAB.5.124202}
Allen, C.~K. and Wangler, T.~P.,
\newblock Phys. Rev. ST Accel. Beams {\bf 5} (2002) 124202.

\bibitem{doi:10.1063/1.37804}
Lysenko, W.~P. and Overley, M.~S.,
\newblock AIP Conference Proceedings {\bf 177} (1988) 323.

\bibitem{Sargsyan_2014}
Sargsyan, K. et~al.,
\newblock International Journal for Uncertainty Quantification {\bf 4} (2014)
  63.

\bibitem{Sargsyan2017}
Sargsyan, K.,
\newblock {\em Surrogate Models for Uncertainty Propagation and Sensitivity
  Analysis}, pages 673--698,
\newblock Springer International Publishing, Cham, 2017.

\bibitem{HandbookOfUQ}
Ghanem, R., Higdon, D., and Owhadi, H., editors,
\newblock {\em Handbook of Uncertainty Quantification},
\newblock Springer International Publishing, Cham, 2017.

\bibitem{Sullivan2015}
Sullivan, T.~J.,
\newblock {\em Introduction to Uncertainty Quantification},
\newblock Springer International Publishing, Cham, 2015.

\bibitem{10.2307/2371268}
Wiener, N.,
\newblock American Journal of Mathematics {\bf 60} (1938) 897.

\bibitem{10.2307/1969178}
Cameron, R.~H. and Martin, W.~T.,
\newblock Annals of Mathematics {\bf 48} (1947) 385.

\bibitem{SpanosGhanem}
Spanos, P.~D. and Ghanem, R.,
\newblock Journal of Engineering Mechanics {\bf 115} (1989) 1035.

\bibitem{doi:10.1137/S1064827501387826}
Xiu, D. and Karniadakis, G.,
\newblock SIAM Journal on Scientific Computing {\bf 24} (2002) 619.

\bibitem{10.2307/2236692}
Rosenblatt, M.,
\newblock The Annals of Mathematical Statistics {\bf 23} (1952) 470.

\bibitem{JAKEMAN2019643}
Jakeman, J.~D., Franzelin, F., Narayan, A., Eldred, M., and Pfl{\"u}ger, D.,
\newblock Computer Methods in Applied Mechanics and Engineering {\bf 351}
  (2019) 643 .

\bibitem{doi:10.1111/j.1467-9868.2005.00503.x}
Zou, H. and Hastie, T.,
\newblock Journal of the Royal Statistical Society: Series B (Statistical
  Methodology) {\bf 67} (2005) 301.

\bibitem{doi:10.1080/00401706.1970.10488634}
Hoerl, A.~E. and Kennard, R.~W.,
\newblock Technometrics {\bf 12} (1970) 55.

\bibitem{doi:10.1111/j.2517-6161.1996.tb02080.x}
Tibshirani, R.,
\newblock Journal of the Royal Statistical Society: Series B (Methodological)
  {\bf 58} (1996) 267.

\bibitem{OMP_Rubinstein}
Rubinstein, R., Zibulevsky, M., and Elad, M.,
\newblock {E}fficient {I}mplementation of the {K-SVD} {A}lgorithm using {B}atch
  {O}rthogonal {M}atching {P}ursuit,
\newblock Technical report, CS Technion, 2008.

\bibitem{NIPS2001_1976}
Figueiredo, M.,
\newblock Adaptive sparseness using jeffreys prior,
\newblock in {\em Advances in Neural Information Processing Systems 14}, edited
  by Dietterich, T.~G., Becker, S., and Ghahramani, Z., pages 697--704, MIT
  Press, 2002.

\bibitem{doi:10.1080/00401706.1991.10484804}
Morris, M.~D.,
\newblock Technometrics {\bf 33} (1991) 161.

\bibitem{GAN2014269}
Gan, Y. et~al.,
\newblock Environmental Modelling \& Software {\bf 51} (2014) 269 .

\bibitem{Sobol1990}
Sobol', I.~M.,
\newblock Matematicheskoe Modelirovanie {\bf 2} (1990) 112,
\newblock (in Russian), English version in: Mathematical Modelling and
  Computational Experiments, 1(4):407-414, 1993.

\bibitem{HOMMA19961}
Homma, T. and Saltelli, A.,
\newblock Reliability Engineering \& System Safety {\bf 52} (1996) 1 .

\bibitem{efron1979}
Efron, B.,
\newblock Ann. Statist. {\bf 7} (1979) 1.

\bibitem{MARELLI201867}
Marelli, S. and Sudret, B.,
\newblock Structural Safety {\bf 75} (2018) 67 .

\bibitem{doi:10.1080/00949659708811825}
Archer, G. E.~B., Saltelli, A., and Sobol', I.~M.,
\newblock Journal of Statistical Computation and Simulation {\bf 58} (1997) 99.

\bibitem{PhysRevAccelBeams.22.064602}
Frey, M., Snuverink, J., Baumgarten, C., and Adelmann, A.,
\newblock Phys. Rev. Accel. Beams {\bf 22} (2019) 064602.

\bibitem{KOLANO201854}
Kolano, A., Adelmann, A., Barlow, R., and Baumgarten, C.,
\newblock Nuclear Instruments and Methods in Physics Research Section A:
  Accelerators, Spectrometers, Detectors and Associated Equipment {\bf 885}
  (2018) 54 .

\end{thebibliography}

\end{document}